\documentstyle[12pt]{article}

\newlength{\extraspace}
\newlength{\extraspaces}
\newcommand{\be}{\begin{equation}}
\newcommand{\ee}{\end{equation}}
\newcommand{\bear}{\begin{eqnarray}}
\newcommand{\eear}{\end{eqnarray}}
\newcommand{\ba}{\begin{array}}
\newcommand{\ea}{\end{array}}
\newcommand{\li}{\mbox{${\rm Li}_{2}\/$}}
\newcommand{\acosh}{\mbox{${\rm arccosh}$}}
\newcommand{\asinh}{\mbox{${\rm arcsinh}$}}

\newcommand{\trace}{\mbox{\rm Tr}}

\newcommand{\rtoo}{\mbox{$\sqrt{2}$}}
\newcommand{\pauli}{\mbox{$\vec{\tau}$}}
\newcommand{\ww}{\mbox{$W_{L}W_{L}$}}

\newcommand{\vw}{\mbox{$\vec{w}$}}
\newcommand{\vpse}{\mbox{$\vec{p}$}}
\newcommand{\vpi}{\mbox{$\vec{\pi}$}}
\newcommand{\spr}{\mbox{$\sigma^{\prime}$}}
\newcommand{\ppr}{\mbox{$\vec{\pi}^{\prime}$}}
\newcommand{\fs}{\mbox{$\Phi^\prime$}}
\newcommand{\fso}{\mbox{$\fs_{0}$}}
\newcommand{\szero}{\mbox{$\Sigma_{0}$}}
\newcommand{\Dmu}{\mbox{$D^\mu$}}
\newcommand{\dmu}{\mbox{$D_\mu$}}
\newcommand{\pdmu}{\mbox{$\partial_\mu$}}
\newcommand{\PDmu}{\mbox{$\partial^{\mu}$}}
\newcommand{\su}{\mbox{$SU(2)\times U(1)$}}
\newcommand{\spin}{\mbox{$Sp\,(4)$}}
\newcommand{\stwo}{\mbox{$SU(2)\times SU(2)$}}
\newcommand{\lr}{\mbox{$SU(2)_{L}\times SU(2)_{R}$}}
\newcommand{\ew}{\mbox{$SU(2)_{W}\times U(1)_{Y}$}}
\newcommand{\hc}{{\rm h.c.}}
\newcommand{\gev}{\mbox{\rm GeV}}
\newcommand{\wz}{\mbox{$w^{+} w^{-} \rightarrow z z$}}
\newcommand{\lpr}{\mbox{$\lambda^{\prime}$}}
\newcommand{\yu}{\mbox{\rm U}}
\newcommand{\xee}{\mbox{$\xi$}}
\newcommand{\xpr}{\mbox{$\xi^{\prime}$}}
\newcommand{\xdpr}{\mbox{$\xi^{\prime\prime}$}}
\newcommand{\ctri}{\mbox{$\lambda_{3}$}}
\newcommand{\ctes}{\mbox{$\lambda_{4}$}}
\newcommand{\cA}{{\cal A}} \newcommand{\cN}{{\cal N}}
\newcommand{\cB}{{\cal B}} \newcommand{\cO}{{\cal O}}

 \newcommand{\cT}{{\cal T}}
 \newcommand{\cU}{{\cal U}}
\newcommand{\cI}{{\cal I}} \newcommand{\cV}{{\cal V}}
 
 \newcommand{\cX}{{\cal X}}
\newcommand{\cL}{{\cal L}} 
\newcommand{\cM}{{\cal M}} 
\newcommand{\np}{Nucl.\ Phys.\ {\bf B}}
\newcommand{\pr}{Phys.\ Rev.\ }
\newcommand{\prd}{Phys.\ Rev.\ {\bf D}}
\newcommand{\prp}{Phys.\ Rep.\ }
\newcommand{\prl}{Phys.\ Rev.\ Lett.\ }
\newcommand{\pl}{Phys.\ Lett.\ {\bf B}}
\newcommand{\jsp}{J.\ Stat.\ Phys.\ }
\newcommand{\cmp}{Comm.\ Math.\ Phys.\ }

\newcommand{\ap}{Ann.\ Phys.\ }
\newcommand{\rmp}{Rev.\ Mod.\ Phys.\ }

\begin{document}
\pagestyle{empty}

\begin{titlepage}
\begin{flushright}
{\rm BUHEP-93-30 \\ hep-ph/9312317 \\ December 20, 1993}
\end{flushright}

\vspace{24pt}
\begin{center}
{\LARGE The Phenomenology of a Non-Standard  \\[0.2cm]
 Higgs Boson in $W_{L}W_{L}$ Scattering}\\
\vspace{40pt}
{\rm Vassilis Koulovassilopoulos}\footnote{e-mail:vk@budoe.bu.edu},
{\rm R. Sekhar Chivukula}\footnote{e-mail:sekhar@abel.bu.edu}
\vspace*{0.5cm}

{\it Physics Department, Boston University,\\
590 Commonwealth Avenue,\\
Boston, MA 02215 USA}
\vskip 1.0cm
\rm
\vspace{25pt}
{\bf ABSTRACT}
\vspace{12pt}
\end{center}
\baselineskip=18pt
\begin{minipage}{15.5cm}

In this paper we consider the phenomenology of a ``non-standard'' Higgs
Boson in longitudinal gauge-Boson scattering. First, we present a
composite Higgs model (based on an $SU(4)/Sp\,(4)$ chiral-symmetry
breaking pattern) in which there is a non-standard Higgs Boson. Then we
explore, in a model-independent way, the phenomenology of such a
non-standard Higgs by calculating the chiral logarithmic corrections to
longitudinal gauge scattering. This calculation is done using the
Equivalence theorem and the Higgs is treated as a scalar-isoscalar
resonance coupled to the Goldstone Bosons of the \mbox{$SU(2)_{L} \times
SU(2)_{R}/SU(2)_{V}$} chiral symmetry breaking. We show that the most
important deviation from the one Higgs-doublet standard model is
parameterized by one unknown coefficient which is related to the Higgs
width. The implications for future hadron colliders are discussed.
\end{minipage}

\vfill
\end{titlepage}

\pagebreak
\baselineskip=18pt
\pagestyle{plain}
\setcounter{page}{1}

\section{Introduction}
\hspace*{\parskip}
Although it is well established that the electroweak interactions are
described by a spontaneously broken \su\ gauge theory, the
underlying physics of the symmetry breaking is still elusive.
In the minimal (one Higgs-doublet) Standard Model (SM), electroweak
symmetry breaking is assumed to be due to an \lr\ Gell-Mann-Levy
linear sigma model. In the broken phase this theory has a neutral
scalar particle, the Higgs  Boson ($H$), which, along with the
isotriplet ($w^a$) of Goldstone Bosons of the spontaneously broken
\lr\ symmetry (which become the longitudinal components of the $W^\pm$
and $Z$) completes a complex scalar doublet.  The sigma-model dynamics
provide the Higgs a mass which is proportional to the quartic
self-coupling $\lambda$, and is unconstrained by any symmetry.  If the
Higgs is heavy, the model becomes strongly coupled and perturbation
theory breaks down \cite{kn:thacker}.

However, there are good reasons to believe that no scalar field theory
can be  fundamental. First, scalar field theories are ``unnatural''
\cite{kn:thooft}: since no (ordinary) symmetry protects scalar masses,
theories without supersymmetry require a large amount of fine tuning to
maintain a hierarchy between the weak scale and any higher scale in the
theory. There is, however, a stronger restriction on such theories coming
 from the analysis of their short distance behavior. Consideration
of the dynamics seems to suggest that scalar field theories are trivial in
the continuum limit \cite{kn:phi4,kn:o(n)}: for a physically
meaningful bare coupling ($\lambda_0$) the renormalized coupling ($\lambda_R$)
is forced to be zero in the limit that the cutoff ($\Lambda$) is sent to
infinity. As the cut-off decreases, the upper bound on $\lambda_R$, and,
consequently, on the Higgs mass $m_H=2 \lambda_R v^2$, increases. Of course,
in order that the theory make sense, $m_H$ must be less than $\Lambda$.
This sets a maximum upper bound, which is estimated to be roughly $600-800$
GeV, if $\Lambda$ is of order a few TeV \cite{kn:dn}. This suggests that
the standard model can only be viewed as a low energy effective theory
below some scale $\Lambda$ where additional new physics enters.

Physically, it is more relevant to interpret the triviality bound by
turning the argument around: if the electroweak symmetry breaking
(EWSB) sector involves a heavy (iso-)scalar resonance that couples to
the electroweak gauge Bosons, then one should expect that it has
properties rather different from those of the SM Higgs and that these
deviations become larger as the mass of this putative Higgs
grows. Such a particle we generically call a ``non-standard Higgs
Boson''. With luck, the physics of symmetry breaking will be directly
probed \cite{kn:ehlq} in the next generation of high energy colliders
(LHC), perhaps through longitudinal gauge-Boson scattering
\cite{kn:cahn}.

In this paper we will compute the chiral-logarithmic corrections to
longitudinal gauge-Boson scattering in a theory with a non-standard Higgs
Boson. At sufficiently high energy, using the Equivalence theorem \cite
{kn:thacker,kn:chanowitz,kn:ET}, the (strong) scattering amplitudes of
the Goldstone Bosons which would be present in the absence of the
electroweak gauge symmetry are approximately the same as those of the
longitudinal electroweak gauge  Bosons. The interactions of Goldstone
Bosons are conveniently described in the language of chiral Lagrangians
\cite{kn:ccwz,kn:gm,kn:bible}. To lowest order in momentum, the
most general effective theory which contains an isoscalar ``Higgs''
and the isotriplet of Goldstone Bosons of the spontaneously broken
\lr\ symmetry (at energies below the cut-off scale $\Lambda$)
\cite{kn:ccwz,kn:short} is
\be
\cL = \frac{1}{4} (v^2 + 2 \xee v H + \xpr H^2 + \xdpr
\frac{H^3}{6 v} )\; \trace\, (\pdmu \yu^{\dagger}\PDmu\yu) \;\;  + \;\;
\cL_{H}  \label{eq:efl}
\ee
where \xee\ , \xpr\, and \xdpr\ are unknown coefficients, $v=246 \ {\rm
GeV}$, and \yu\ contains the Goldstone Bosons $w^a$\
\be
\yu = \exp \left(\frac{i \vw \cdot \pauli}{v} \right) \;\;\;\; , \;\;\;
       \trace \,  (\tau^{a} \tau^{b}) = 2 \delta^{a b}
\ee
with \pauli\ the Pauli matrices and $\cL_{H}$\ be the Lagrangian for the
isoscalar
\be
\cL_{H} = \frac{1}{2} (\pdmu H)^2 -  \frac{m^2}{2} H^2 -
    \frac{\ctri v}{3 !}H^3 - \frac{\ctes}{4 !}H^4 \label{eq:Lh}.
\ee
Here $SU_L (2)$ corresponds to $SU(2)_{\it weak}$ while the $\tau_3$
component of the custodial $SU(2)_R$ corresponds to hypercharge.
In addition, for simplicity, we have included only the leading
non-derivative terms in the scalar potential (which are the only ones
relevant for this investigation).  The ordinary linear sigma-model
corresponds to the limit
\be
  \xee \, , \, \xpr \; = \; 1
   \;\;\; , \;\;\;   \xdpr \; = \; 0  \label{eq:newone}
\ee
and
\be
  \ctri\, , \, \ctes \; = \; \frac{3 m^2}{v^2} \ \ . \label{eq:newtwo}
\ee

In the next section, we will illustrate the possibility of a non-standard
Higgs by constructing a Composite Higgs model of the Georgi-Kaplan type
\cite{kn:gk,kn:dgk} based on an $SU(4)/Sp\,(4)$ symmetry breaking pattern.
Here the compositeness scale will be identified with the cut-off $\Lambda$
of the Composite Higgs effective theory. In the limit that $\Lambda
\rightarrow \infty$, any Composite Higgs model reduces to the Standard
Model. However, if the Higgs is heavy, new physics must enter at a scale of
the order a few TeV and deviations from the Standard Model may be large
\cite{kn:short}. For this example, we derive the relationship between the
parameters of the Composite Higgs theory and the parameters in the
Lagrangian eq.~({\ref{eq:efl}) above.

In the third section, we present the calculation of the
chiral-logarithmic corrections to longitudinal gauge Boson scattering
in a model-independent manner by starting from the general effective
theory of eq.~(\ref{eq:efl}).  This section elaborates on the results
presented in \cite{kn:short}. The second and third sections of the
paper are essentially independent of one another.

The fourth section contains our conclusions.  Finally, Appendix A
contains an explicit form of the Lagrangian for the $SU(4)/Sp(4)$
model, Appendix B contains the Feynman rules for the Lagrangian
(\ref{eq:efl}), and Appendix C the analytical expressions for the
one-loop integrals used in section 3.

\section{The $SU(4)/Sp(4)$ model}
\label{sec-su4sp4}
\hspace*{\parskip}
In this section we shall describe an explicit model which contains a
non-standard Higgs Boson. We focus here only on the phenomenology of the
longitudinal gauge-Boson scattering. The ordinary fermions won't enter
our analysis and, therefore, we neglect the issue of ordinary fermion
mass generation (though it is straightforward to extend the model to
generate fermion masses).

The model assumes the existence of a new strong and confining
``ultracolor'' interaction based on a gauge group $G_{c}$, four new
left-handed {\em ultrafermions} transforming in a pseudo-real
representation of $G_{c}$ (recall that there must be an even number of
fermions in order to avoid the Witten anomaly
\cite{kn:witten}), and a fundamental scalar doublet. The presence of a
fundamental scalar doublet is somewhat unsatisfactory. However, as we will
see, the mass of the non-standard Higgs in this model can be as large as
700 GeV without the self-interactions of the fundamental scalar doublet
being particularly strong: in this sense the theory is {\it less
trivial} than the usual fundamental scalar-doublet standard model with a
heavy Higgs.

In the limit that the ultrafermions are massless and the limit of
vanishing Yukawa coupling between scalar and the fermions and the
fundamental scalar, the global chiral symmetry of the ultrafermions is
$G =SU(4)$. These fermions have electroweak quantum numbers: one
$SU(2)_W$ doublet with hypercharge $Y = 0$\ and two $SU(2)_W$ singlets
with $Y =\pm 1/2$. The explicit form of the \ew\ generators embedded
in the flavor group $G$ is given by
\be
\vec{S} = \frac{1}{2} \left( \begin{array}{cc}
      \pauli & 0 \\  0 & 0    \end{array} \right)  \;\;\; , \;\;\;
   Y =  \frac{1}{2} \left( \ba{cc}   0 & 0 \\ 0 &  \tau_{3}
                \ea  \right)           \label{eq:ewgenerators}
\ee
where \pauli\ are the Pauli matrices and as usual the electromagnetic
charge is generated by $Q = S_3 +Y$.

When the ultracolor interactions $G_{c}$ become strong,
at the ``chiral symmetry breaking'' scale $\Lambda$ (which will
be of order a TeV or higher), a condensate is produced
\be
\langle \psi \psi^{\top} \rangle \approx \Lambda f^2\, \Delta \label{eq:cond}
\ee
where $\top$ denotes the transpose in ultracolor space, $f$ is the
``f-constant'' for ultracolor chiral symmetry breaking (the analog of
$f_{\pi}=93$ MeV in QCD), and $\Delta$ is a unitary matrix in flavor space
that characterizes the vacuum orientation. The rules of ``Naive Dimensional
Analysis'' \cite{kn:gm} imply that $\Lambda$ must be less than or of order
$4 \pi f$.

By making a $G$ transformation, $\Delta$\ can be brought to the form
\be
\Delta = \left( \ba{cc} \tau_{2} & 0 \\ 0 & \tau_{2} \ea \right)\ \ .
               \label{eq:delta}
\ee
The condensate (\ref{eq:cond}) spontaneously breaks the chiral symmetry
$G$ down to the subgroup $H = Sp\,(4)$
\cite{kn:preskill,kn:kosower,kn:leshouches},
producing five Goldstone  Bosons (which would be massless in the absence of
electroweak gauge interactions, and the Yukawa couplings and fermion mass
terms described below). Notice that \ew\ is contained in $H =
Sp\,(4)$ and thus is not broken at the scale $\Lambda$ by the ultrafemion
condensate. Furthermore, electroweak radiative corrections will not induce
such a breaking \cite{kn:preskill}. As we show in the next section, the
Yukawa couplings to the fundamental scalar will be responsible for
misaligning the vacuum slightly and driving electroweak symmetry breaking.

The $10$ unbroken \spin\ generators $\cN^i$\ and the broken ones $\cX^a$
satisfy the relations\footnote{The first relation follows from the
definition of the $Sp\,(4)$ algebra.}
 \be
 \Delta \cN^\top \, \Delta = - \cN \;\;\; , \;\;\;
 \Delta \cX^\top \, \Delta = \cX\;\;.           \label{eq:genid}
 \ee
The broken generators $\cX^a$, conveniently
normalized as $\trace\, (\cX^a \cX^b) =\frac{1}{2}\delta^{ab}$, are
given by
\be
X = \frac{1}{2\sqrt{2}}\left( \ba{cc} 0 & I \\ I & 0 \ea \right)
      \; , \;\;
\vec{T} = \frac{1}{2\sqrt{2}}\left( \ba{cc} 0 & - \pauli \\ \pauli & 0
 \ea \right) \; \; , \;\;
A = \frac{1}{2\sqrt{2}}\left( \ba{cr} I & 0 \\ 0 & -I \ea \right)\ \ .
\ee
The algebra of $Sp\,(4)$ is isomorphic to that of
$SO\,(5)$, which contains an $SO\,(4) \simeq \stwo$ subgroup  given
by eq.~(\ref{eq:ewgenerators}), with $Y$ being the third component of the
custodial $SU(2)$ symmetry.  Thus the five Goldstone  Bosons
$\Pi^{\alpha}=(\sigma, \vec{\pi},a) \; ,
\alpha =1, \ldots ,5$, fall into a representation which
decomposes under \stwo\ as the sum of a {\bf (2,2)}
\be
\Phi = \sigma X  + i \, \vec{\pi} \cdot \vec{T}
\ee
that has the quantum numbers of the usual Higgs doublet, and a {\bf
(1,1)}, the $a$, that appears as an  electroweak singlet.
For convenience we will denote the complex doublet of fundamental scalars
\fs , which also transforms under \lr\ as $\bf (2,2)$, by
\be
\fs = \spr X  + i \, \ppr \cdot \vec{T}\ .
\ee

Below the chiral symmetry breaking scale $\Lambda$, the
dynamics of the Goldstone  Bosons can be described by an effective chiral
Lagrangian, in analogy with QCD \cite{kn:ccwz,kn:gm,kn:bible}. Using
a nonlinear realization, the $\Pi^\alpha$ are incorporated
into the field $\xi$
\be
\xi = \exp \left( \frac{ i\, \Pi^\alpha \cX^\alpha}{f}\right)
\; , \;\;\;\;\;\;\; \xi\,\rightarrow \,\xi^\prime\, =\, g\,\xi\, h^\dagger
\ee
where $g \in G,\; h \in H$. As usual, it is more convenient to define
the field $\Sigma= \xi^2$ which transforms linearly under $H$. By using
eq.~(\ref{eq:genid}) and noting that  $\xi^2 = \xi \Delta \xi^{\top}
\Delta$, it is easy to show that $\Sigma$ transforms under $G$ like
\bear
\Sigma & \rightarrow & g \, \Sigma \, g^{\dagger} \;\;\; , \;\;\;\;
g  \in H
 \nonumber \\
\Sigma & \rightarrow & g \, \Sigma \, g \;\;\;\; , \;\;\;\;
g \in G/H\;\;.
\eear
The interactions of the $\Pi$ with the \ew\ gauge  Bosons are
described to lowest order in momentum by a gauged $G/H$ nonlinear
sigma-model
\be
\cL_\Sigma = \frac{f^2}{4} \trace\, (\dmu\Sigma^\dagger\,\Dmu\Sigma )
 + \frac{M^2 f^2}{2} \trace\, (\Sigma + \Sigma^\dagger)
                \label{eq:lsigma}
\ee
with the covariant derivative defined as
\be
\dmu\Sigma = \pdmu\Sigma + i g [\vec{S},\Sigma]\!\cdot\!\vec{W}_\mu
     +  i g^\prime [Y,\Sigma]\; \cB_{\mu}\;\;.
\ee
In the second term in eq.~(\ref{eq:lsigma})
we have assumed that the $\Pi^{\alpha}$ have a mass
due to a $G_c$ and \ew\ invariant mass term for the ultrafermions. In
the fundamental theory, this fermion mass term is of the form: $\cM \psi^{\top}
\Delta \psi$. In the low-energy theory $M^2 = \mu \cM$,
where $\mu$ is a dimensionful constant of order $\Lambda$.

The interactions of the fundamental scalar \fs\ are described by the usual
$\phi^4$ Lagrangian, given by
\be
\cL_{\phi} =\trace\, (\dmu\fs^\dagger \Dmu\fs) -\lambda \left[ \trace
 \,(\fs^\dagger \fs)+ \frac{m_{\phi}^2}{2 \lambda} \right]^2 .
        \label{eq:lfs}
\ee
Here $m^{2}_{\phi}>0$, so that the electroweak symmetry is not broken
through the interactions in eq.~(\ref{eq:lfs}) alone.

Finally, \fs\ and $\Pi$ mix through Yukawa couplings of \fs\ with the
ultrafermions $\psi^{i}$. Above the chiral symmetry breaking scale
$\Lambda$ this interaction is of the
form $iy\, \psi^{\top} \Delta \fs \psi\, + \,\hc$. In the effective theory,
to lowest order in $y$, it becomes an interaction between
\fs\ and $\Pi$ given by
\be
{\cal L}_{Yuk} = \frac{i y \mu f^2}{2} \,\trace \, \fs\,
( \Sigma^{\dagger} - \Sigma ) . \label{eq:yuk}
\ee
Notice that in eq.~(\ref{eq:yuk}) we have included only pseudo-scalar
couplings. In principle we could  also have scalar couplings, but for the
vacuum to align so as to break the electroweak symmetry it is essential
that there are non-zero pseudo-scalar couplings. The addition of
scalar couplings of the fundamental doublet would not qualitatively
change the analysis.

Now we can gather all the pieces together to write the Lagrangian
$\cL_{eff}$ of the $SU(4)/Sp\,(4)$ model for scales below $\Lambda$. To
lowest order in momentum and the other symmetry breaking terms ($M$ and
$y$), it is given by
\be
\cL_{eff} =  \cL_\Sigma + \cL_\phi + \cL_{Yuk} . \label{eq:mlag}
\ee

\subsection{Vacuum Alignment}
\hspace*{\parskip}
Let us now examine the ground state of the Lagrangian in
eq.~(\ref{eq:mlag}). Since we do not want to break electromagnetism,
we search for a minimum of the potential of our effective theory with
expectation values only for the two iso-singlets $(\sigma , \spr)$.  (All
other vacua which preserve electromagnetism are equivalent to these up
to a chiral rotation.)  Setting $\vec{\pi}=\ppr=a=0$, the potential is
given by
\bear
\lefteqn{ V(\sigma , \spr) = \frac{m_{\phi}^2}{2} \spr^{2} +
\frac{\lambda}{4} \spr^{4}\; - }   \nonumber \\*[0.1cm] & &
   4 M^2 f^{2} \cos\!\left(\frac{\sigma}{\sqrt{2} f}\right) -
 \sqrt{2}\mu f^2 y\,\spr\sin\!\left(\frac{\sigma}{\sqrt{2}f}\right).
                       \label{eq:vpot}
\eear
The conditions $\partial V/ \partial
\sigma =0 \mbox{  and  } \partial V/ \partial \spr = 0$, which
determine the extrema, always have
the trivial solution $\sigma=\spr=0$, for arbitrary values of
the couplings.
However, a solution exists for non-zero values of $\sigma$\ and \spr\ if
\be
\frac{yf\mu}{\sqrt{2} M m_{\phi}} > 1 \;\;\; , \; \;\;  m_{\phi} > 0
\ee
\noindent
and this solution has lower energy than $\sigma=\spr=0$.
That is, for $y > y_c = \sqrt{2} M m_{\phi}/\mu f$ the vacuum
becomes unstable, giving vev's to both $\sigma$ and \spr, and
thereby breaking \ew. One combination of the the \vpi\
and \ppr\ become the longitudinal components of $W^{\pm}$ and $Z$, while
the orthogonal combinations remain in the spectrum as a degenerate (due
to isospin) pseudo-scalar isotriplet. Expanding eqs.~(\ref{eq:lsigma}),
(\ref{eq:lfs}) and requiring that the $W$ and $Z$ masses
be correct, we obtain
\be
 v^{2} = 2 f^{2} \sin^{2}\frac{\langle \sigma \rangle}{\rtoo f}
    + \langle  \spr  \rangle^{2}   \label{eq:weakscale}
\ee
where $v=246$ GeV.

It is convenient to recast our description in the more
familiar notation of the two-Higgs doublet model by defining
\bear
 \sqrt{2} f \sin \frac{\langle \sigma \rangle}{\rtoo f}  & = &
  v \cos\!\alpha \nonumber \\*[0.2cm]
  \langle \spr \rangle  & = & v \sin\!\alpha\ \ .   \label{eq:aangle}
\eear
In order to identify the mass eigenstate fields, we perform a field
redefinition in $\cL_{eff}$ and rewrite $\Sigma$\ and \fs\  as
\be
\Sigma = \cU^{\dagger} \, \szero \, \cU \;\;\;\;\;\;  , \;\;\;\;\;\;\;\;
\fs = \cV^{\dagger} \, \fso \,\cV
\ee
where
\be
\cU = \exp \left(\frac{2 i \vec{S} \cdot \vpi}{v \cos\!\alpha }\right)
      \;\;\;\;\;\;   , \;\;\;\;\;\;
\cV = \exp \left(\frac{2 i \vec{S} \cdot \ppr}{v \sin\!\alpha }\right)
  \label{eq:parametr}
\ee
and
\bear
 \szero & = & \exp\left[ \frac{2 i X}{f} (\sigma +
        \langle \sigma \rangle )\right]
   \nonumber \\[0.3cm]
  \fso & = &   (\spr + \langle \spr \rangle) X
\eear
where for simplicity $\sigma \mbox{ and }\spr $\ denote now the
shifted fields. This field redefinition leaves all S-matrix elements
invariant \cite{kn:salam}, and the resulting Lagrangian obtained after some
straightforward but tedious algebra is provided in eq.~(\ref{eq:susp}) in
appendix A.

It is now straightforward to identify the mass eigenstates \vw\ and
\vpse\ as those obtained from \vpi\ and \ppr\ by making a rotation by
$\alpha$
\be
   \left( \ba{c} \vw \\ \vpse \ea \right)  =
   \left( \ba{rc} \cos\alpha   & \sin\alpha  \\
               -\sin\alpha   &  \cos\alpha \ea \right)
   \left( \ba{c} \vpi \\ \ppr \ea \right) .
\ee
The \vw\ correspond to the exact Goldstone Bosons that are eaten by the
longitudinal states of $W_{a}^{\mu}$, while \vpse\ remains in the
spectrum as a degenerate massive isotriplet. 

The $2\times 2$ mass-squared matrix for the two neutral
(iso-singlet) scalar states $\sigma$\ and \spr, defined by
\be
 \left. \frac{\partial^{2} V}{\partial \sigma_{i}
     \partial \sigma_{j}} \right|_{ \sigma=0}
\ee
is obtained from eq.~(\ref{eq:vpot}) by using
eqs.~(\ref{eq:weakscale}) and
(\ref{eq:aangle}). Its explicit form is given by
\be
  y \mu f \left( \ba{cc}
  \tan\alpha &         \parbox{1.5in}{
  \[\mbox{}- \sqrt{1- \frac{v^2 \cos^2 \!\alpha}{2 f^2}} \] }
    \\*[-0.4cm] \parbox{1.5in}{
  \[\mbox{}- \sqrt{1- \frac{v^2 \cos^{2}\!\alpha}{2 f^2}}\]  }
       &    \parbox{1.5in}{
  \[ \frac{1}{\tan \alpha} + \frac{2 \lambda v^2 \sin^2 \!\alpha}{y f
  \mu}  \]  }  \ea \right) . \label{eq:hHmass}
\ee
The eigenstates of this symmetric matrix correspond to the
iso-singlet mass eigenstates $H,h$, and are found by performing a
rotation\footnote{In the literature of the two-Higgs
doublet models the angles $\alpha \mbox{  and  } \beta$\ are
sometimes defined the other way around, namely $\alpha$ rotating
the iso-singlet states and $\beta$\ rotating the iso-triplets.}
on $\sigma$\ and \spr
\be
\left( \ba{c} H \\ h \ea \right) =
 \left( \ba{rc} \cos \beta  & \sin \beta \\
         -\sin \beta & \cos \beta \ea \right)
  \left( \ba{c} \sigma  \\ \spr \ea \right)\ \ .
\ee
The expressions for the masses $m , m_h $ of $H$ and $h$ respectively,
as well as the angle $\beta$ are not particularly illuminating except in
some limiting cases that we discuss below.

Two ranges of the parameters are of particular interest: (i) Having $f
\gg v$ and small Yukawa coupling and (ii) having $f$\ of $\cO\/(v)$ and
Yukawa coupling $y \simeq \cO\/(1)$. In the first case, by taking the
scale $f \gg v$ we find that $\alpha -\beta \sim\cO (v^2 /f^2)$. The
masses of the heavy Higgs $h$, the massive isotriplet \vpse\ and the
singlet $a$, grow
with $f$, with the $h$ and $\vpse$ becoming degenerate. All of these
states decouple from light states $H , w^a $, in the usual sense of
Appelquist and Carazzone \cite{kn:cara}. On the other hand, the mass of
the ``light'' Higgs $m$ does not grow with $f$. Setting $\mu \approx 4
\pi f$, in the $f \rightarrow \infty$ limit $m$ tends to
\be
m^2 \longrightarrow  2 \,
(\lambda \sin^4 \alpha +  \pi y \sin\alpha \cos^3 \alpha)\, v^2\ \ .
\ee
In this limit\footnote{Of course this limit requires a large amount of fine
tuning in order to maintain the hierarchy between $f$ and $v$.}
the $SU(4)/Sp\/(4)$ model reduces exactly to the Standard
model in the spontaneously broken phase written in ``polar'' coordinates
\[
\sigma + v + i\pauli \cdot \vpi\; \equiv\; (H+v)\; \exp
\left(\frac{i \vw \cdot \pauli}{v}\right)\ \ ,
\]
with the value of the $\lambda_{\phi^4}$ coupling renormalized at scale
$\Lambda$ equal to $\lambda \sin^4 \alpha + \pi y \sin\alpha \cos^3
\alpha$.

Case (ii), where $f \simeq \cO\/(v)$, however, is more interesting
since it is the regime where the ``light'' scalar Higgs is entirely
``non-standard''. In order to give a feeling for this regime, we
take as reference the values
\be
f =180 \gev \;\; , \;\; \lambda = 1 \;\; , \;\; \alpha =
30^{\circ} \;\; , \;\; y = 3.5  \label{eq:modval}
\ee
For these values the mass spectrum is
\bear
  m = 718 \; \gev \;\; , &  m_{h}= 1675 \; \gev  \;\; , &
  m_{p}=1814 \gev \; , \nonumber \\ m_a= 907 \;\gev   \; ,
    &  \beta = 21.5^{\circ} \;\; ,  &  \Lambda \approx 4 \pi f
\simeq 2.2 \; \mbox{\rm TeV}\ \ .    \label{eq:values}
\eear
As advertised, the Higgs mass is of order 700 GeV without the
self-coupling of the fundamental scalar being particularly large.
The Lagrangian eq.~(\ref{eq:mlag}) does include only the {\it lowest
order} terms in the effective chiral theory. However, using
the rules of dimensional
analysis \cite{kn:gm} we see that, for the values chosen in
eq.~(\ref{eq:modval}), the higher order terms should not significantly
change the vacuum structure of the theory \cite{kn:gk}.

In both cases and, indeed, for generic values of the parameters this model
has a mass gap between the lowest lying states $H,w^a$ and the other heavy
particles $h, p^a ,a$. Hence it is reasonable to focus on the interactions
of the light scalars. ``Integrating out'' the heavy states, the
$SU(4)/Sp\/(4)$ theory has precisely the form proposed in
eq.~(\ref{eq:efl}), in terms of a single isoscalar resonance with the
values of $\xi ,\xpr ,\xdpr ,\ctri ,\ctes$ given as functions of $f,
\lambda , \alpha , y$ in Appendix A. Generically we have
\be
  \xee \, , \, \xpr \; = \; 1 + \cO \, \left(\frac{v^2}{f^2}\right)
   \;\;\; , \;\;\;   \xdpr \; = \;\cO\,\left( \frac{v^2}{f^2} \right)
\ee
and
\be
  \ctri\, , \, \ctes \; = \; \frac{3 m^2}{v^2} + \,
   \cO \, \left(\frac{v^2}{f^2}\right)\ \ .            \label{eq:lsml}
\ee
In particular, for the reference values in eq.~(\ref{eq:modval})
\be
\xi=0.62\;\; ,\; \xpr=-0.21\;\; ,\;\xdpr=0.71\;\; ,\;
\ctri= 18.26\;\; ,\; \ctes= 4.79\ \ .      \label{eq:vl}
\ee

\section{Longitudinal Gauge  Boson scattering}
\label{sec-wlwl}
\hspace*{\parskip}
Using the Equivalence theorem
\cite{kn:thacker,kn:chanowitz,kn:ET}, high energy \ww\ scattering
amplitudes in the full gauge theory can be reliably computed purely in
the scalar sector by replacing the longitudinal components of the
gauge Bosons with the corresponding Goldstone Boson. In this section,
we will analyze the Lagrangian in eq.~(\ref{eq:efl}) which describes,
to lowest order in momentum, the most general theory with a massive
scalar-isoscalar resonance coupled to the isotriplet of exact
Goldstone Bosons of the \lr\ symmetry spontaneously broken down to
$SU(2)_{L+R}$. In particular, we explore the consequences of such a
``non-standard Higgs'' Boson for longitudinal gauge Boson
scattering. The discussion given here completes the analysis begun in
ref.~\cite{kn:short}.

Chiral Perturbation theory \cite{kn:gm,kn:bible,kn:gl} has been used
extensively to study the phenomenology of gauge-Boson scattering, but
generally under the assumption that the {\it only}\footnote{An
interesting alternative that includes  potentially light scalars is
technicolor models with additional fundamental scalars \cite{kn:samuel}.
However, in contrast to the $SU(4)/Sp\,(4)$ model presented above, these
do not reduce to the standard model by taking any particular limit.}
strongly interacting degrees of freedom lighter than a TeV were the
longitudinal gauge-Bosons themselves
\cite{kn:golden,kn:dobado,kn:donoghue}. If there were
a Higgs as heavy as a TeV, then for energies lower than that it would
be sufficient to ``integrate out'' the Higgs and estimate its effects
at lower energies on the processes involving only $W^\pm ,Z$
\cite{kn:donoghue,kn:ecker}. However, here the Higgs resonance is
light enough compared to the cut-off $\Lambda$ or the other heavy
resonances in the model, so we must include it explicitly in the
effective Lagrangian. Note that, in this limit, the theory is {\it
very} different than QCD, in which there is no significant mass gap
between the lightest resonance and any of the others.

We present a calculation of the chiral logarithmic corrections to
longitudinal gauge  Boson scattering \wz. (All other channels
can be obtained from this one by crossing.) The treatment of infinities
induced by loops follows the standard rules of effective nonrenormalizable
theories \cite{kn:gm,kn:gl}. Namely, the infinities associated with
non-derivative interactions in the Lagrangian eq.~(\ref{eq:efl}) are
absorbed in the renormalization of the scalar self-couplings, while those
associated with vertices involving derivatives are absorbed in \lr\
invariant counterterms which are of order $p^4$. The coefficients of these
counterterms are new independent (running) couplings in the theory and
cannot be computed. In what follows, we shall compute the leading
corrections in the $\overline{MS}$ scheme, setting the higher order
counterterms to zero when the renormalization scale $\mu$ is taken to be
$\Lambda \simeq 4 \pi f$. These results include the so-called ``chiral
logarithms'', which are the leading contributions at sufficiently low
momentum \cite{kn:pagels}. As required by consistency of the
chiral expansion \cite{kn:gm,kn:bible}, the chiral logarithms are
expected to be comparable in magnitude to the $p^4$\ counterterms.

In addition, in our computation we have kept the ``finite'' parts of
$\cO\/(p^4)$\ terms that come along with the logarithms in the
$\overline{MS}$ scheme. Unlike the chiral logarithms, these corrections
are scheme-dependent and therefore, in general they have no physical
meaning. However, in the limit $f \rightarrow \infty$ the theory reduces to
the one Higgs-doublet standard model and, as in any renormalizable theory,
all of the $\mu$-dependence of our answers should disappear. In this limit
we should recover the exact one-loop amplitude for the linear sigma model
provided in ref.~\cite{kn:dawil}. This will serve as an important check for
our calculation.

The Feynman rules for the Lagrangian eq.~(\ref{eq:efl}) are given in
Appendix B.

\subsection{Tree amplitude}

The tree amplitude for \wz\ is found from the diagrams in Fig.~1 to be
\be
\cA_{tree} = \frac{s}{v^2} - \left(\frac{\xi^2}{v^2}\right)
 \frac{s^2}{s-m^2} . \label{eq:tree}
\ee
where $s,t,u$\ are the usual Mandelstam variables with $s+t+u=0$.
For $\sqrt{s} \ll m,v$\ only the first term (coming solely from
the GB contact diagram) survives; this corresponds to  the low energy theorems
\cite{kn:chanowitz,kn:golden}. For somewhat higher energies deviations from
the SM emerge and these are parameterized by just one parameter \xee. For
$\xi \neq 1$, the tree amplitude in eq.~(\ref{eq:tree}) has a
bad high energy behavior, as expected for a nonrenormalizable effective
theory. In fact, for $\xi < 1$ the amplitude vanishes for some energy
greater than $m^2$. However, in this region, the tree amplitude is not
trustworthy since higher order effects will be large. Furthermore, in the
region around the peak $s=m^2$ we have to include the Higgs decay width.
This is done through the substitution
\be
\frac{i}{s -m^2} \longrightarrow \frac{i}{s-m^2 -i \mbox{\rm
  Im}\Pi_{H}^{(1)}(s)  }
 \label{eq:trpro}
\ee
where  $\Pi_{H}^{(1)}(s)$ is the one-loop self-energy given in
eq.~(\ref{eq:self}) below\footnote{The inclusion of the running width
instead of the constant one has been suggested \cite{kn:valencia92} to
be more effective on peak.}.


\subsection{One-loop amplitude}

We now present the one-loop amplitudes which are formally of order
$\cO\/(p^{4})$ in the chiral expansion. These are calculated using
$\overline{MS}$ regularization with the infinities in the diagrams
(which correspond to poles in $1/\epsilon$) omitted. Hence the
parameters $m,v,\xee,\xpr,\xdpr,\ctri,\ctes$\ that
appear in the expressions below are actually the $\overline{MS}$\
renormalized quantities. The corresponding physical Higgs mass and VEV,
denoted here as $m_{H}, \nu$, are obtained from $m,v$\ by a finite
renormalization\footnote{Actually $m_{H}$ is the on-shell mass and not
the real part of the pole position.}
\bear
\nu^2 & = & Z_{v}^{-1} v^2        \nonumber\\*[0.2cm]
  m_{H}^{2} & = &  m^2 + \delta\, m^2           \label{eq:physcon}
\eear
with $\nu = 246\; \gev$ and
\be
\delta m^{2} =  \mbox{\rm Re}\,\Pi_{H}^{(1)}(m_{H}^{2})
\ee
with $\Pi_{H}^{(1)}(p^2)$ the one-loop Higgs self-energy given by
eq.~(\ref{eq:self}) below, with $G(1)= -2 + \pi/\sqrt{3}$.
We  used above that $Z_{v}=Z_{w}$, the wavefunction renormalization of
the $w^a$. This is found either directly, by minimizing  the 1-loop
effective potential and determining the shift from $v$, or by using the Ward
identity that ensures the Equivalence theorem\footnote{See
eq.~(4.11)-(4.13) in the paper by Bagger and Schmidt in
ref.~\cite{kn:ET} and also \cite{kn:durand}.}.
Let us remark finally that the slightly unconventional choice of the vev
$v$ as a running quantity is substituted in the more traditional
treatments of electroweak radiative corrections by the mass of the gauge
 Boson $M_W$, since $M_W = g v/2$. For our purpose the differences
between various definitions of $g$ are negligible \cite{kn:marciano}.

\noindent
{\bf A. Self-Energy Contributions}

The Higgs self-energy $\Pi_{H}^{(1)} (p^{2})$\ is found from the diagrams of
Fig.~2a  to be
\be
\Pi_{H}^{(1)}(s)= -\frac{1}{16\pi^2} \left\{ \frac{3 \xi^2 s^2}{2 v^2}
 B_{1}(s) + \frac{m^{2}\ctes}{2} A(m) - \frac{\ctri^2 v^2}{2}
 \left[1+G\left(\frac{s}{m^2}\right)\right] \right\}. \label{eq:self}
\ee
Analytical expressions for the functions used are given in Appendix C.
The imaginary part of the one-loop self-energy $\Pi_{H}^{(1)}$\ is
related to  the decay width through a unitarity cut
\be
\mbox{\rm Im}\Pi_{H}^{(1)}(m^{2}) = - m \Gamma_{H}^{(0)} \label{eq:cut}
\ee
and is given by
\be
\Gamma_{H}^{(0)} = \frac{3 m^3}{32 \pi v^2} \xi^2 . \label{eq:trwid}
\ee
Since the parameter \xee\ can only be $\xi < 1$ in the $SU(4)/Sp\,(4)$
model, we see that in this model such a Higgs would be narrower than the SM
Higgs \footnote{In models with elementary scalars, since no scalar VEV can
exceed 246 GeV, $\xi$ must be less than one. In general we know of
no reason that this must be the case, although we know of no model that has
$\xi > 1$.}.

The one-loop contribution of $\Pi_{H}^{(1)}(s)$ to \wz, for
energies away from the Higgs-Boson pole, is given by
\be
\cA_{2-pt} = - \left(\frac{\xi^2}{v^2}\right)
\frac{s^{2}\;\Pi_{H}^{(1)}(s)}{(s-m^2 )^{2}}
\ee
In the resonance region, $|s-m^2| \leq m \Gamma_{H}$, again we have to
include the Higgs decay in the propagator. However  now, if we want to
count powers of $\lpr( \equiv m^{2}/2 v^{2})$ (which in the linear sigma
model is the self-coupling $\lambda$) consistently  on the peak, we have
to re-sum not only the imaginary part of the one-loop Higgs self energy
but also its real part along with the imaginary part of the two-loop
Higgs self energy $\Pi_{H}^{(2)}(s)$. Furthermore, since we are using a
non-linear representation of the Goldstone Bosons, the inclusion of the
width is consistent with the Equivalence theorem \cite{kn:valencia}.
Hence, the Higgs self-energy contribution on peak to \wz, is found by
making the replacement (in order to avoid double counting) of the
diagram in Fig.~1b, which corresponds to the second term in
eq.~(\ref{eq:tree}), with the diagram in Fig.~2c with the propagator
given by
\be
\frac{i}{s -m^2} \longrightarrow \frac{i}{s-m^2 - \Pi_{H}^{(1)}(s)
     -i\mbox{\rm Im}\Pi_{H}^{(2)}(s)}     \label{eq:pro}
\ee
With this re-summation procedure our complete amplitude is correct to
order $\cO\,(\lpr)$\ on the peak (see \cite{kn:shape}). The imaginary
part of $\Pi_{H}^{(2)}(s)$ at $s=m^2$ can be found from the one-loop
correction, $\Gamma_{H}^{(1)}$, to the Higgs decay width through a
unitarity cut similar to that of eq.~(\ref{eq:cut}) at the appropriate
order. This is presented in the following subsection.

 From $\Pi_{H}$\ in eq.~(\ref{eq:self}), and the GB self-energy
$\Pi_{w}$, given by the diagrams in Fig.~3, we find  the
wavefunction renormalization factors, defined as
\begin{eqnarray*}
Z_{w} & = & 1 + \left. \mbox{\rm Re}\frac{d\Pi_{w}}{d\,
p^{2}}\right|_{p^{2}=0} \\*[0.2cm]
Z_{H} & = & 1 + \left. \mbox{\rm Re}\frac{d\Pi_{H}}{d\,
p^{2}}\right|_{p^{2}=m^{2}}
\end{eqnarray*}
to be given by
\bear
Z_{w}^{1/2} & = & 1 - \frac{m^{2}}{32 \pi^{2}v^{2}} \left\{
  \frac{\xi^2}{2} + \left[ \xi^2 - \xpr + \frac{\xi\ctri v^2}{m^2}
    \right]   A(m) \right\} \\*[0.2cm]
Z_{H}^{1/2} & = & 1 - \frac{1}{32 \pi^{2}} \left\{
  \frac{3 \xi^2 m^2}{v^2}\left[\frac{1}{2}+ A(m)\right] - \frac{\ctri^2
      v^2}{2 m^2}  \left(1 -\frac{2 \pi \sqrt{3}}{9} \right)  \right\}.
\eear

\noindent
{\bf B. Vertex Corrections}

The $H w^{+} w^{-}$\ and $H z z$\  three-point functions contribute to
\wz\ scattering through the diagrams shown in Fig.~4. The $H w^{+}
w^{-}$\ three-point function is
\bear
i\, \Gamma (s) & = & \frac{i}{16 \pi^2 v^3} \left\{ \mbox{}
   -  \xi\xpr m^2 s - \xi s^2 B_{1}(s)\; -\; \frac{\xpr\ctri v^2}{2} s
    \left[1+ G\left(\frac{s}{m^2}\right) \right]   \right.
 \nonumber \\*    
& & - \left(\xdpr  + \xi\xpr \right) m^2 s A(m)\;  +
     \xi^{3} s  \left[ \frac{m^2}{2} + \left(
       m^2 - \frac{s}{2} \right) B_1(s) - m^2 C_{1}(s) \right]
  \nonumber \\*    
& &  \left. \mbox{} + \xi^2 \ctri v^2 \left[ m^2 A(m) -
         \left(\frac{s}{2} + m^2 \right )  B_{2}(s)  -
          m^2 C_{2}(s) \right] \right\}.
\eear
Isospin invariance implies that the $H z z$\ vertex is identical to the
$Hw^{+}w^{-}$\ vertex. This contribution to the \wz\ amplitude is
\be
\cA_{3-pt} =  \left(\frac{\xi}{v}\right) \,
\frac{2\; s \; \Gamma (s)}{s-m^2}
\ee
where the replacement in eq.~(\ref{eq:pro}) is needed in the resonance
region.

 From the proper three-point function we can compute the decay width of
the Higgs at one-loop, by adding to the tree amplitude the real part of
$\Gamma (s)$ (for $s=m^2$), and multiplying each external $H, w^a$\
line by wavefunction renormalization factors $Z_{H}^{1/2}, Z_{w}^{1/2}$,
respectively. This gives
\be
\left\{ -\frac{\xi m_{H}^{2}}{\nu} +\mbox{\rm Re}\,
\Gamma\,(m_{H}) \right\} Z_{w}^{1/2} Z_{H}^{1/2}  \label{eq:width}
\ee
where we have used eq.~(\ref{eq:physcon}) to replace $v$\ with $\nu =
246\;\gev$\ in the tree amplitude.
Then the one-loop correction $\Gamma_{H}^{(1)}$ to the total Higgs decay
width $\Gamma_{H}^{1-loop}=\Gamma_{H}^{(0)} + \Gamma_{H}^{(1)}$, is
found to be (see also \cite{kn:short})
\bear
\lefteqn{\frac{\Gamma_{H}^{(1)}}{\Gamma_{H}^{(0)}} = \frac{1}{8\pi^2}
  \left\{ \frac{m^2}{v^2}\left[1+A(m)\right]+\frac{\xpr\ctri}{2\xi}\left[
  \frac{\pi}{\sqrt{3}}-1 \right] + \frac{\ctri^2 v^2}{4 m^2}
   \left[ 1-\frac{2\pi\sqrt{3}}{9}\right]  +\frac{\xdpr
      m^2}{\xi v^2} A(m) \right. } \nonumber\\ & &
\left. \mbox{}+ \xpr \frac{3m^{2}}{2 v^{2}} \left[\frac{1}{3} +A(m)
       \right] +  \frac{\xi^2 m^2}{2v^2} \left[\frac{\pi^2}{6}
  -4- A(m)\right]- \frac{\xi\ctri}{2}\left[\pi\sqrt{3}-3-\frac{2\pi^2}{9}
  \right]   \right\} \label{eq:hwidth}
\eear
where we used the explicit form of the $B,C$ functions presented in
Appendix C. In the linear sigma model limit eqs.~(\ref{eq:newone}),
(\ref{eq:newtwo}) our calculation agrees with that in
ref.~\cite{kn:marciano,kn:dawil}. For our reference values of
eq.~(\ref{eq:vl}), the one-loop correction to the tree-level width in
eq.~(\ref{eq:trwid}) is $28 \%$, while the corresponding
standard model correction is only $7.6 \%$.

\noindent
{\bf C. 1-PI Corrections}

The 1-PI diagrams that contribute are shown in Fig.~5. The bubble
diagrams in Fig.~5a  give the chiral logarithmic corrections from the
non-linear sigma model. They are given by
\bear
\lefteqn{\cA_{GB} =  \frac{1}{(4\pi v^2)^{2}} \left\{
   \frac{5}{9} s^{2} +\frac{13}{18} (t^{2} +u^{2})\; +
            \right.} \nonumber \\*[0.1cm]
& &  \left.\;\;\;\; \frac{s^{2}}{2}\, \ln\frac{\mu^{2}}{-s} +
       \frac{t}{6} (s+2 t)\,\ln \frac{\mu^{2}}{-t} +
        \frac{u}{6} (s+2 u)\,\ln\frac{\mu^{2}}{-u}  \right\}
\eear
The rest of the diagrams with two internal propagator given in Fig.~5b
are
\be
\cA_{bubble}=\frac{1}{(4\pi v^2)^2} \left\{ \frac{\xpr^2}{2} s^2
    B_{2}(s) + 2 \xi^2 m^2 s \left[\frac{1}{2} + A(m)\right] \right\}
\ee
The diagrams with three internal propagator shown in Fig.~5c are given
by
\bear
\lefteqn{\cA_{1-tri} = \frac{2 \xi^{2}}{3 (4\pi v^2)^{2}} \left\{
  \frac{2 s^{2}-t^{2}-u^{2}}{6}\; - 3 m^2 s\;  + 6 m^{2} \left[ m^2 +
       \frac{s}{2} - \frac{m^2 s^2}{t u} \right] A(m) \right. }
      \nonumber\\*[0.1cm]
& & \mbox{} +3s \left(\frac{s}{2}-m^{2}\right) B_{1}(s) - \;
 \left[ 3m^{4}+t^{2}+\frac{st}{2}-\frac{3}{2} m^{2}t+6 m^{4}\frac{s}{t}
 \right] B_{1}(t)   \nonumber\\*[0.1cm]
& & \mbox{} - \left[ 3m^{4}+ u^{2}+\frac{su}{2}-\frac{3}{2} m^{2}u
  +6 m^{4}\frac{s}{u}  \right] B_{1}(u) + 3 m^{2} s C_{1}(s)
          \nonumber\\*[0.1cm]
& & \mbox{} +\left. 3 m^2 \left[ s+m^2 +2 m^2\frac{s}{t} \right]
     C_{1}(t)+ 3 m^2 \left[ s+m^2 +2 m^2 \frac{s}{u} \right] C_{1}(u)
    \right\}
\eear
while the rest with three internal propagator in Fig.~5d  are
\be
\cA_{2-tri} = \frac{2 \xi^{2}\,\xpr}{(4\pi v^2)^{2}} \left\{ m^{2}s A(m)
   - \left(\frac{s^2}{2}+m^2 s \right) B_{2}(s) - m^2 s C_{2}(s) \right\}
\ee
The diagrams in Fig.~5e, with four internal propagator are equal to
\be
\cA_{box} = \cI(s,t)\; + \;\cI(s,u)
\ee
where
\bear
\lefteqn{\cI (s,t) = \frac{\xi^4}{(4\pi v^2)^2}
\left\{ \frac{s^2 + t^2 + 4 s t}{18}+ \frac{m^2}{3}(2 s + t)
 \right. }        \nonumber\\[0.1cm]
& & \mbox{}
  -  m^2 \left[ 6 m^2 +\frac{1}{3}(2 s-5t)+ 4m^2 \left(\frac{s}{t} +
\frac{t}{s} \right) \right] A(m)   \nonumber\\[0.1cm]
& &
\mbox{}+ \left[ 3 m^4 + 4m^4 \frac{s}{t} - m^{2}t +\frac{t}{6}(s+2 t)
  \right] B_{1}(t)   \nonumber\\[0.1cm]
& & \mbox{}+ \left[ 3 m^{4}+\frac{2}{3}m^{2}(s-t) + 4m^{4}\frac{t}{s} +
  \frac{s}{6} (2 s +t) \right] B_{2}(s)
\nonumber\\[0.1cm]  & &  \mbox{}+
2 m^2 \left[ 2m^2 -t+ 2m^2 \frac{t}{s} \right] C_{2}(s)-
2 m^2 \left[ 2 m^2 + s + 2 m^2 \frac{s}{t} \right] C_{1}(t)
\nonumber\\[0.1cm]  & &  \left.
 \mbox{} +  m^4 D(s,t)  \right\}.
\eear

\noindent
{\bf D. Tadpoles}

The contribution from the tadpole diagrams shown in Fig.~6 is
\be
\cA_{tad}=\frac{1}{(4\pi v^2)^{2}}\left\{ \xi\ctri v^{2}- \xpr
            m^{2}\right\} s A(m).
\ee

\noindent
{\bf E. Complete one-loop amplitude}

Adding all the one-loop contributions from above, and multiplying by
a factor of $Z_{w}^{1/2}$ for each external line we find that the
complete one-loop correction to the \wz\ amplitude is  given by
\bear
\cA_{1-loop} & =  & \cA_{2-pt}+ \cA_{3-pt} +\cA_{GB} +\cA_{bubble}
+\cA_{1-tri} + \cA_{2-tri}\nonumber\\*[0.1cm]
& & +\cA_{box} + \cA_{tad}
+ 4 \cA_{tree}(Z_{w}^{1/2} - 1) \label{eq:lamp}
\eear
As explained, our results here expressed in terms of the renormalized
mass and VEV. The physical parameters are found through the relations
(\ref{eq:physcon}). We have verified explicitly that, in the linear
sigma model limit eqs.~(\ref{eq:newone}-\ref{eq:newtwo}), our
computation is scale independent and reproduces the calculation of
ref.~\cite{kn:dawil}. This provides a non-trivial check of our
calculation.

For the sake of completeness we present here the low energy limit of our
amplitude which constitute the $p^4$ corrections to the low energy
theorems \cite{kn:short}. For this we use the low energy limit of the
the functions provided in Appendix C. These can be found, for example,
in Appendix~A in ref.~\cite{kn:dawsontop}. Let us call the total
amplitude for \wz, up to one-loop, by
\be
\cA (s,t,u)= \cA_{tree} + \cA_{1-loop}\ \ .
\ee
In the limit  $s \ll m^2 \; \mbox{\rm and}\; -t \ll m^2$, we find
\be
\cA(s,t,u) = \frac{s}{v^2} + \frac{1}{(4\pi v^{2})^{2}}\; \cT
 + \xi^2  \frac{s^2}{m^2 v^2}  \label{eq:lowexp}
\ee
\bear
\cT & = & \frac{s^2}{2} \ln\frac{\mu^2 }{-s} + \frac{t}{6} (s +2t)
     \ln \frac{\mu^2}{-t} + \frac{u}{6} (s+ 2u)\ln\frac{\mu^2}{-u}
\nonumber\\*[0.1 cm]  & & \mbox{} +
   s^2 \, P \; + \; Q\, (t^2 +u^2 ) \; + \, R\ln \frac{m^{2}}{\mu^2}
\eear
where
\bear
 P & = & \frac{5}{9} + 2  \xi \xdpr  + \xi^2 \left( \frac{7}{2} \xpr +
     \frac{22}{9} \right)  - \frac{65}{9} \xi^4
   + \frac{\xi \ctri}{2 \lambda^\prime} \left(\xpr - \frac{\xi^2}{2}\right)
  + \xi^2  \frac{\ctri^2}{8 {\lambda^\prime}^{2}} \left(
      \frac{\pi}{\sqrt{3}} - 2  \right) 
\\[0.1 cm]
Q  & = &  \frac{13}{18} -  \frac{11}{9}\xi^{2} + \frac{5}{18} \xi^{4}
\\[0.1 cm]
R  & = & s^{2}\, \left[ \frac{37}{6}\xi^4 - \xi^2 \left( \frac{10}{3}+ 2
           \xpr \right) +2 \xi \xdpr - \frac{\xpr^2}{2}  \right]
  + \frac{\xi^2}{3} \left( 2 - \xi^2 \right) \left(t^2 +u^2 \right)
\eear
where $\lambda^\prime= m^2/2 v^2$.
Also in eq.~(\ref{eq:lowexp}), and since $m < 4 \pi v$, we have retained the
$1/m^2$ correction to this order in the momentum expansion.  In the
linear sigma model limit and to leading order in $s/m^2$ this amplitude
explicitly agrees with that of ref.~\cite{kn:dawil}.

The scattering amplitudes for all other channels can be obtained
 from the \wz\ amplitude $\cA(s,t,u)$.
In terms of the kinematical variables $s,t,u$, these are given by
\bear
 \cA (\wz) & = & \cA(s,t,u) \nonumber\\*[0.1cm]
 \cA (w^{+}w^{-} \rightarrow w^{+}w^{-}) & = & \cA(s,t,u) + \cA(t,s,u)
\nonumber\\*[0.1cm]
 \cA (z z \rightarrow z z) & = & \cA(s,t,u)+\cA(t,s,u)+\cA(u,t,s)
\nonumber\\*[0.1cm]
 \cA (w^{+} w^{+} \rightarrow w^{\pm} w^{\pm}) & = & \cA(t,s,u)+ \cA(u,t,s)
\nonumber\\*[0.1cm]
 \cA (w^{\pm} z \rightarrow w^{\pm} z) &=& \cA(t,s,u)
\eear

\subsection{Cross Section and Discussion}

The differential cross section for longitudinal gauge  Boson scattering
in any of these channels, is obtained from the amplitudes above by
\be
\frac{d\sigma}{dt}=\frac{1}{16 \pi s^{2}}\left| \cA \right|^{2}
\ee
where $\cA =  \cA_{tree} +\cA_{loop}$.
Since we neglected the 1-PI two-loop diagrams, we have
\be
| \cA |^{2} = |\cA_{tree}|^{2} + 2\left\{\, \mbox{\rm
Re}\,( \cA_{tree})\mbox{\rm Re}\,(\cA_{loop})+ \mbox{\rm
Im}\,(\cA_{tree})\mbox{\rm Im}\,(\cA_{loop})\, \right\}\ \ .
\ee
The total cross section is
\be
\sigma_{tot}\/(s)=\int_{-s}^{0}dt\,\frac{d\sigma}{dt}(s,t)
\ee

In Fig.~7a we show the total cross section for the $W_L^+ W_L^-
\rightarrow W_L^+ W_L^-$\ channel as a function of $s$, for the
$SU(4)/Sp\,(4)$ model with the parameters in eq.~(\ref{eq:vl}). The
height of the peak for the one-loop curve is only $4.5 \%$ higher than the
peak in the tree-level curve.  The corresponding curves for a Standard
model Higgs with the same mass are shown in Fig.~7b; here the one-loop
peak exceeds the tree-level peak by $11 \%$.  The sharp fall in the
amplitude in the region above the peak can be understood by noticing
that for $\xi < 1$\ the tree amplitude in eq.~(\ref{eq:tree}) vanishes
at some energy greater than $m^2$ (if one does not include a finite
width). This only signals that higher order effects are expected to be
significant there. Also, far above the peak the amplitude presented is
not trustworthy due to the breakdown of the expansion in powers of
$1/\Lambda$.

Qualitatively, however, for gauge-boson scattering below a TeV, the
width and shape of the peak appear to be the most important features
differentiating a standard from a non-standard Higgs resonance.  In
Fig.~8a we show the decay width $\Gamma_{H}$\ as a function of mass
for the non-linear model for our reference values in eq.(\ref{eq:vl}),
both at tree-level and one-loop, while the corresponding graph for the
Standard model is shown in Fig.~8b.

The cross sections discussed above are not directly measurable in
hadron colliders like the LHC; one must first convolute them with the
\ww\ luminosities inside the proton. A more detailed study of how well the
LHC be able to differentiate a standard from a non-standard Higgs can
only be answered after detailed analysis of a specific detector. This
question is currently under investigation. Unlike the analysis
presented here, however, one must also include the potentially
non-standard couplings of this non-standard Higgs to the top-quark
since this affects the production of the Higgs through gluon fusion.

\section{Conclusions}

In this paper, we considered the phenomenology of a ``non-standard''
Higgs Boson. In the first half of our analysis, we discussed a
composite Higgs model based on an $SU(4)/Sp\,(4)$\ non-linear sigma
model that features such a non-standard Higgs. This model illustrates
the possibility of the existence of a scalar resonance that is
relatively light (possibly within the reach of the LHC), but whose
dynamics is significantly different from that of the Standard Model
Higgs Boson. Unlike conventional technicolor models, composite Higgs
models may have a large mass gap between the Higgs and the other
heavier resonances and, therefore, one cannot necessarily rely on the
discovery of the other, more exotic, resonances as an experimental
signature of the dynamics. Instead, it will be important to understand
whether the light scalar resonance has the properties of a Standard
model Higgs or not.

Furthermore, because of the triviality of fundamental scalar theories,
{\it any} scalar Boson with a mass of order 500 GeV or higher which
couples to the longitudinal electroweak gauge Bosons is likely to have
properties very different than those of the Standard Model Higgs.

In the second half of our analysis we calculated the chiral
logarithmic corrections to longitudinal gauge-Boson scattering in a
theory with a non-standard Higgs resonance. We found that the most
important deviations from the Standard Model are
parameterized by the parameter \xee, which is directly related to the
Higgs width.

The understanding of the electroweak symmetry sector will surely require
further experimental investigation. If a scalar iso-scalar resonance is
observed in longitudinal gauge-Boson scattering, it will be important
to show whether or not it is the Standard Model Higgs Boson.

\section{Acknowledgments}
We would like to thank Mitchell Golden, Dimitris Kominis Stephen
Selipsky, and Elizabeth Simmons for helpful discussions and
suggestions. V.K. also thanks Ulrich Nierste for checking the form of
the box diagram given in Appendix C.  R.S.C.  acknowledges the support
of an Alfred P. Sloan Foundation Fellowship, an NSF Presidential Young
Investigator Award, a DOE Outstanding Junior Investigator Award, and a
Superconducting Super Collider National Fellowship from the Texas
National Research Laboratory Commission.  This work was supported in
part under NSF contract PHY-9057173 and DOE contract
DE-FG02-91ER40676, and by funds from the Texas National Research
Laboratory Commission under grant RGFY92B6.

\section*{Appendix A}

In this appendix we give the expression for the Lagrangian
of the $SU(4)/Sp\/(4)$\ that we mentioned in section~\ref{sec-su4sp4}.
Making the field redefinition given by eq.~(\ref{eq:parametr}), we have
\be
\szero = \cos\left[ \frac{\sigma +\langle\sigma\rangle}{\rtoo f}\right]
  + i 2 \sqrt{2} X \sin\left[\frac{\sigma
            +\langle\sigma\rangle}{\rtoo f}\right]
\ee
and the $SU(4)/Sp\/(4)$\ Lagrangian eq.~(\ref{eq:mlag}) is
\bear
 \lefteqn{\cL_{f} = \frac{1}{2} \left( \PDmu\sigma\pdmu\sigma \right) +
 \frac{f^2}{2}\sin^{2} \left[\frac{\sigma +\langle\sigma\rangle}{\rtoo
  f} \right] \trace\left( \PDmu\cU\pdmu\cU^{\dagger}\right) -  }
\nonumber \\*[0.1cm]  & &  V(\sigma,\spr) +
 \frac{1}{2}\PDmu\spr\pdmu\spr +\frac{1}{4} \left( \frac{\spr
 +\langle\spr\rangle}{\rtoo f} \right)^{2}
\trace \left( \PDmu\cV\pdmu\cV^{\dagger} \right)
 \nonumber \\*[0.1cm]  & &
  - \frac{\rtoo f m^{2}_{p}}{v^{2}\sin 2 \alpha} \left(\spr
  + \langle\spr\rangle \right) \sin\left[\frac{\sigma
 +\langle\sigma\rangle}{\rtoo f}\right] \;  \vpse\cdot\vpse
                           \label{eq:susp}
\eear
where $V(\sigma,\spr)$\ is the potential in eq.~(\ref{eq:vpot}),
expressed in terms of the shifted fields $\sigma, \spr$.
Above we have ignored the singlet field $a$, and terms with more
pseudo-scalars \vpse, coming from $\cL_{Yuk}$, since these won't be
relevant to our calculation. The \vpse\ and $a$ masses are
\be
m^{2}_{p} = \frac{2 y \mu f}{\sin 2\alpha} \;\;\;\; ,
\;\;\;\;\;\;\; m^{2}_{a} =  y \mu f \tan\!\alpha
\ee

In the Lagrangian above, there are some individual diagrams, like the $H
p^{a}p^{a}$\ vertex coming from the $\cL_{Yuk}$, that grow like
$f^{2}/v$. Such terms would appear to violate the decoupling of the
\vpse\ as $f \rightarrow \infty$. However before the redefinition of
eq.~(\ref{eq:parametr}), there are no terms that grow with
$f$. The resolution to this dilemma lies in a general theorem in
Quantum Field Theory \cite{kn:ccwz,kn:salam}, according to which all
physical on-shell S-matrix elements are invariant under field
redefinitions which leave unchanged the one-particle states, which
constitute the Hilbert space of the theory. Indeed, one can easily
verify that, when evaluating any S-matrix element, the bad large-$f$
behavior cancels when the external lines go on-shell.

Finally, after ``integrating-out'' the heavy states \vpse, $h$ and $a$,
we obtain the Lagrangian in eq.~(\ref{eq:efl}). The parameters \xee,
\xpr, \xdpr, \ctri, \ctes\  are given by
\bear
\xee & = & \sin\alpha \sin\beta + \cos\alpha \cos\beta
            \sqrt{1-\frac{v^{2} \cos^{2} \alpha}{2 f^{2}}}
            \nonumber \\[0.1cm]
\xpr & = & 1- \frac{v^{2}}{f^{2}} \cos^{2}\alpha \cos^{2}\beta
              \nonumber \\[0.1cm]
\xdpr & = & \frac{v^{2}}{f^{2}}\cos\alpha\cos^{3}\beta\;
          \sqrt{1-\frac{v^{2}\cos^{2}\alpha}{2 f^{2}}}
             \nonumber \\[0.1cm]
\ctri & = & 6v\; (\lambda \sin\alpha \sin^{3}\beta + \pi y \cos\alpha
           \sin\beta \cos^{2}\beta )   \nonumber \\[0.1cm]
\ctes & = & 6\lambda \sin^{4}\beta + 2\pi y\cos^{3}\beta \left\{4\sin\beta
             \sqrt{1-\frac{v^{2}\cos^{2}\alpha}{2 f^{2}}}
             -\tan\alpha \cos\beta \right\} .
\eear

\newpage

\input FEYNMAN

\section*{Appendix B}
In this appendix we give the Feynman rules for the Lagrangian in
eq.~(\ref{eq:efl}).


\begin{picture}(40000,55000) 
\drawline\fermion[\E\REG](5000,52000)[2000]
\drawarrow[\E\ATTIP](\pbackx,\pbacky)
\put(4000,\pbacky){$p$}

\drawline\scalar[\E\REG](1500,51000)[4]
\put(19000,51600){\makebox(0,0)[l]{\begin{minipage}{1.3in}
    \[ \frac{i}{p^{2}-m^{2}} \]     \end{minipage}}}
\put(0,\pbacky){$H$}
\put(11000,\pbacky){$H$}
\drawline\fermion[\E\REG](1500,45000)[8500]
\put(19000,45600){\makebox(0,0)[l]{\begin{minipage}{1.3in}
    \[ \frac{i}{p^{2}} \]     \end{minipage}}}
\put(0,\pbacky){$w^{a}$}
\put(11000,\pbacky){$w^{a}$}
\drawline\fermion[\E\REG](5000,46000)[2000]
\drawarrow[\E\ATTIP](\pbackx,\pbacky)
\put(4000,\pbacky){$p$}
\drawline\fermion[\E\REG](1500,37000)[4500]
\drawline\scalar[\NE\REG](\fermionbackx,\fermionbacky)[3]
\drawline\scalar[\SE\REG](\fermionbackx,\fermionbacky)[3]
\put(0,\fermionbacky){$H$}
\drawline\fermion[\NE\REG](7500,39700)[2000]
\drawarrow[\NE\ATTIP](\pbackx,\pbacky)
\put(6500,39000){$p_{1}$}
\drawarrow[\NE\ATTIP](\pbackx,\pbacky)
\drawline\fermion[\SE\REG](7500,34200)[2000]
\put(6500,34600){$p_{2}$}
\drawarrow[\SE\ATTIP](\pbackx,\pbacky)
\put(11000,41000){$w^{+},z$}
\put(11000,32000){$w^{-},z$}
\put(19000,37600){\makebox(0,0)[l]{\begin{minipage}{1.3in}
   \[- i\xi \frac{2p_{1}\cdot  p_{2}}{v} \] \end{minipage}}}
\drawline\fermion[\SE\REG](1500,28000)[5600]
\put(0,\pfronty){$H$}
\drawline\scalar[\NE\REG](\pbackx,\pbacky)[3]
\put(11000,\pbacky){$w^{+},z$}
\drawline\scalar[\SE\REG](\pfrontx,\pfronty)[3]
\put(11000,\pbacky){$w^{-},z$}
\drawline\fermion[\SW\REG](\fermionbackx,\fermionbacky)[5600]
\put(0,\pbacky){$H$}
\drawline\fermion[\NE\REG](7500,27200)[2000]
\drawarrow[\NE\ATTIP](\pbackx,\pbacky)
\put(6500,26500){$p_{1}$}
\drawline\fermion[\SE\REG](7500,21000)[2000]
\put(6500,21000){$p_{2}$}
\drawarrow[\LDIR\ATTIP](\pbackx,\pbacky)
\put(19000,24600){\makebox(0,0)[l]{\begin{minipage}{1.3in}
  \[-i\xi^{\prime}\; \frac{2p_{1}\cdot p_{2}}{v^{2}}\] \end{minipage}}}
\drawline\fermion[\SE\REG](1500,15000)[5600]
\put(0,\pfronty){$H$}
\drawline\scalar[\NE\REG](\pbackx,\pbacky)[3]
\put(11000,\pbacky){$w^{+},z$}
\drawline\scalar[\SE\REG](\pfrontx,\pfronty)[3]
\put(11000,\pbacky){$w^{-},z$}
\drawline\fermion[\SW\REG](\fermionbackx,\fermionbacky)[5600]
\put(0,\pbacky){$H$}
\drawline\fermion[\W\REG](\scalarfrontx,\scalarfronty)[4000]
\put(0,\pbacky){$H$}
\drawline\fermion[\NE\REG](7500,14500)[2000]
\drawarrow[\LDIR\ATTIP](\pbackx,\pbacky)
\put(6500,13500){$p_{1}$}
\drawline\fermion[\SE\REG](7500,8000)[2000]
\drawarrow[\LDIR\ATTIP](\pbackx,\pbacky)
\put(6500,8000){$p_{2}$}
\put(19000,11600){\makebox(0,0)[l]{\begin{minipage}{1.3in}
  \[i\xi^{\prime\prime}\; \frac{4p_{1}\cdot p_{2}}{v^{3}}\]
                                  \end{minipage}}}
\end{picture}

\newpage

\begin{picture}(44000,60000)
\drawline\fermion[\E\REG](1500,55000)[3700]
\put(0,\pfronty){$H$}
\drawline\fermion[\NE\REG](\pbackx,\pbacky)[5000]
\put(9500,\pbacky){$H$}
\drawline\fermion[\SE\REG](\fermionfrontx,\fermionfronty)[5000]
\put(9500,\pbacky){$H$}
\put(19000,55600){\makebox(0,0)[l]{\begin{minipage}{1.3in}
           \[- i \lambda_{3} v\]      \end{minipage}}}
\drawline\fermion[\SE\REG](1500,47000)[10000]
\put(0,\pfronty){$H$}
\put(9500,\pfronty){$H$}
\drawline\fermion[\NE\REG](1500,\pbacky)[10000]
\put(0,\pfronty){$H$}
\put(9500,\pfronty){$H$}
\put(19000,44100){\makebox(0,0)[l]{\begin{minipage}{1.3in}
           \[- i \lambda_{4} \]      \end{minipage}}}
\drawline\scalar[\SE\REG](1500,35000)[5]
\put(0,\pfronty){$w^{+}$}
\put(9500,\pfronty){$w^{+}$}
\put(3900,\pfronty){$p_{1}$}
\drawline\fermion[\SE\REG](2800,\pfronty)[2000]
\drawarrow[\LDIR\ATTIP](\pbackx,\pbacky)
\drawline\fermion[\NE\REG](6200,\pbacky)[2000]
\drawarrow[\LDIR\ATTIP](\pbackx,\pbacky)
\put(6100,\pbacky){$p_{3}$}
\drawline\scalar[\NE\REG](1500,\scalarbacky)[5]
\put(0,\pfronty){$w^{-}$}
\put(9500,\pfronty){$w^{-}$}
\put(3900,\pfronty){$p_{2}$}
\drawline\fermion[\NE\REG](2800,\pfronty)[2000]
\drawarrow[\LDIR\ATTIP](\pbackx,\pbacky)
\drawline\fermion[\SE\REG](6200,\pbacky)[2000]
\drawarrow[\LDIR\ATTIP](\pbackx,\pbacky)
\put(6100,\pbacky){$p_{4}$}
\put(15000,32000){\makebox(0,0)[l]{\begin{minipage}{3.5in}
    \[ \frac{i}{3 v^{2}}\left[ 2 p_{1}\cdot p_{4} + 2 p_{2}\cdot p_{3} +
 \left( p_{1}- p_{4}\right) \cdot \left( p_{2}- p_{3} \right) \right] \]
                                  \end{minipage}}}
\drawline\scalar[\SE\REG](1500,23000)[5]
\put(0,\pfronty){$w^{+}$}
\put(9500,\pfronty){$z$}
\put(3900,\pfronty){$p_{1}$}
\drawline\fermion[\SE\REG](2800,\pfronty)[2000]
\drawarrow[\LDIR\ATTIP](\pbackx,\pbacky)
\drawline\fermion[\NE\REG](6200,\pbacky)[2000]
\drawarrow[\LDIR\ATTIP](\pbackx,\pbacky)
\put(6100,\pbacky){$p_{3}$}
\drawline\scalar[\NE\REG](1500,\scalarbacky)[5]
\put(0,\pfronty){$w^{-}$}
\put(9500,\pfronty){$z$}
\put(3900,\pfronty){$p_{2}$}
\drawline\fermion[\NE\REG](2800,\pfronty)[2000]
\drawarrow[\LDIR\ATTIP](\pbackx,\pbacky)
\drawline\fermion[\SE\REG](6200,\pbacky)[2000]
\drawarrow[\LDIR\ATTIP](\pbackx,\pbacky)
\put(6100,\pbacky){$p_{4}$}
\put(15000,20000){\makebox(0,0)[l]{\begin{minipage}{3.5in}
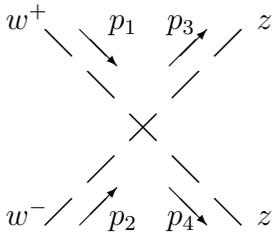
 \[ \frac{i}{3 v^{2}}\left[ 2 p_{1}\cdot p_{2} + 2 p_{3}\cdot p_{4} +
  \left( p_{1}+ p_{2}\right) \cdot \left( p_{3}+ p_{4} \right) \right]
   \]                              \end{minipage}}}
\drawline\scalar[\SE\REG](1500,11000)[5]
\put(0,\pfronty){$w^{+}$}
\put(9500,\pfronty){$w^{+}$}
\put(3900,\pfronty){$p_{1}$}
\drawline\fermion[\SE\REG](2800,\pfronty)[2000]
\drawarrow[\LDIR\ATTIP](\pbackx,\pbacky)
\drawline\fermion[\NE\REG](6200,\pbacky)[2000]
\drawarrow[\LDIR\ATTIP](\pbackx,\pbacky)
\put(6100,\pbacky){$p_{3}$}
\drawline\scalar[\NE\REG](1500,\scalarbacky)[5]
\put(0,\pfronty){$w^{-}$}
\put(9500,\pfronty){$w^{-}$}
\put(3900,\pfronty){$p_{2}$}
\drawline\fermion[\NE\REG](2800,\pfronty)[2000]
\drawarrow[\LDIR\ATTIP](\pbackx,\pbacky)
\drawline\fermion[\SE\REG](6200,\pbacky)[2000]
\drawarrow[\LDIR\ATTIP](\pbackx,\pbacky)
\put(6100,\pbacky){$p_{4}$}
\put(15000,8000){\makebox(0,0)[l]{\begin{minipage}{3.5in}
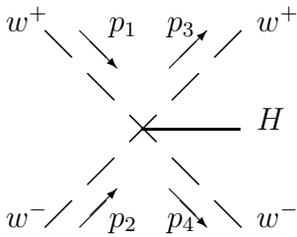
 \[ \frac{2 i\xi}{3 v^{3}}\left[2p_{1}\cdot p_{4}+2p_{2}\cdot p_{3} +
  \left( p_{1}- p_{4}\right)\cdot\left( p_{2}- p_{3} \right)
   \right]    \]                    \end{minipage}}}
\drawline\fermion[\E\REG](5150,7280)[3700]
\put(9500,\pfronty){$H$}
\end{picture}

\pagebreak        

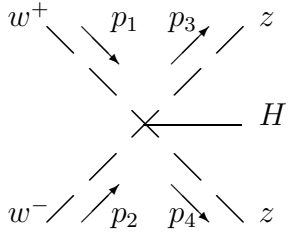
\begin{picture}(44000,50000)
\drawline\scalar[\SE\REG](1500,45000)[5]
\put(0,\pfronty){$w^{+}$}
\put(9500,\pfronty){$z$}
\put(3900,\pfronty){$p_{1}$}
\drawline\fermion[\SE\REG](2800,\pfronty)[2000]
\drawarrow[\LDIR\ATTIP](\pbackx,\pbacky)
\drawline\fermion[\NE\REG](6200,\pbacky)[2000]
\drawarrow[\LDIR\ATTIP](\pbackx,\pbacky)
\put(6100,\pbacky){$p_{3}$}
\drawline\scalar[\NE\REG](1500,\scalarbacky)[5]
\put(0,\pfronty){$w^{-}$}
\put(9500,\pfronty){$z$}
\put(3900,\pfronty){$p_{2}$}
\drawline\fermion[\NE\REG](2800,\pfronty)[2000]
\drawarrow[\LDIR\ATTIP](\pbackx,\pbacky)
\drawline\fermion[\SE\REG](6200,\pbacky)[2000]
\drawarrow[\LDIR\ATTIP](\pbackx,\pbacky)
\put(6100,\pbacky){$p_{4}$}
\put(15000,42000){\makebox(0,0)[l]{\begin{minipage}{3.5in}
 \[ \frac{2 i\xi}{3 v^{3}}\left[2p_{1}\cdot p_{2}+2p_{3}\cdot p_{4} +
  \left( p_{1}+ p_{2}\right)\cdot\left( p_{3}+ p_{4} \right)
   \right]    \]                    \end{minipage}}}
\drawline\fermion[\E\REG](5150,41280)[3700]
\put(9500,\pfronty){$H$}

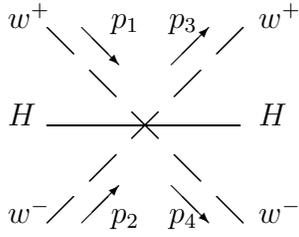
\drawline\scalar[\SE\REG](1500,32000)[5]
\put(0,\pfronty){$w^{+}$}
\put(9500,\pfronty){$w^{+}$}
\put(3900,\pfronty){$p_{1}$}
\drawline\fermion[\SE\REG](2800,\pfronty)[2000]
\drawarrow[\LDIR\ATTIP](\pbackx,\pbacky)
\drawline\fermion[\NE\REG](6200,\pbacky)[2000]
\drawarrow[\LDIR\ATTIP](\pbackx,\pbacky)
\put(6100,\pbacky){$p_{3}$}
\drawline\scalar[\NE\REG](1500,\scalarbacky)[5]
\put(0,\pfronty){$w^{-}$}
\put(9500,\pfronty){$w^{-}$}
\put(3900,\pfronty){$p_{2}$}
\drawline\fermion[\NE\REG](2800,\pfronty)[2000]
\drawarrow[\LDIR\ATTIP](\pbackx,\pbacky)
\drawline\fermion[\SE\REG](6200,\pbacky)[2000]
\drawarrow[\LDIR\ATTIP](\pbackx,\pbacky)
\put(6100,\pbacky){$p_{4}$}
\put(15000,29000){\makebox(0,0)[l]{\begin{minipage}{3.5in}
 \[ \frac{2i\xi^{\prime}}{3 v^{4}}\left[2p_{1}\cdot p_{4}+2p_{2}\cdot
   p_{3} +\left( p_{1}- p_{4}\right)\cdot\left( p_{2}- p_{3} \right)
   \right]    \]                    \end{minipage}}}
\drawline\fermion[\E\REG](1500,28280)[7300]
\put(0,\pfronty){$H$}
\put(9500,\pfronty){$H$}
\drawline\scalar[\SE\REG](1500,19000)[5]
\put(0,\pfronty){$w^{+}$}
\put(9500,\pfronty){$z$}
\put(3900,\pfronty){$p_{1}$}
\drawline\fermion[\SE\REG](2800,\pfronty)[2000]
\drawarrow[\LDIR\ATTIP](\pbackx,\pbacky)
\drawline\fermion[\NE\REG](6200,\pbacky)[2000]
\drawarrow[\LDIR\ATTIP](\pbackx,\pbacky)
\put(6100,\pbacky){$p_{3}$}
\drawline\scalar[\NE\REG](1500,\scalarbacky)[5]
\put(0,\pfronty){$w^{-}$}
\put(9500,\pfronty){$z$}
\put(3900,\pfronty){$p_{2}$}
\drawline\fermion[\NE\REG](2800,\pfronty)[2000]
\drawarrow[\LDIR\ATTIP](\pbackx,\pbacky)
\drawline\fermion[\SE\REG](6200,\pbacky)[2000]
\drawarrow[\LDIR\ATTIP](\pbackx,\pbacky)
\put(6100,\pbacky){$p_{4}$}
\put(15000,16000){\makebox(0,0)[l]{\begin{minipage}{3.5in}
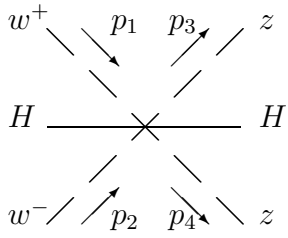
 \[ \frac{2i\xi^{\prime}}{3 v^{4}}\left[2p_{1}\cdot p_{2}+2p_{3}\cdot
   p_{4} +\left( p_{1}+ p_{2}\right)\cdot\left( p_{3}+ p_{4} \right)
   \right]    \]                    \end{minipage}}}
\drawline\fermion[\E\REG](1500,15280)[7300]
\put(0,\pfronty){$H$}
\put(9500,\pfronty){$H$}
\end{picture}

\newpage

\section*{Appendix C}
In this appendix we provide the analytic expressions for the
one-loop integrals that appear in the text. $\mu$\ here, is
the renormalization scale. In our computation we shall take $\mu$
to be the cut-off of our effective theory $\Lambda \simeq 4\pi f$.
We use the notation
\be
\sigma = s/m^2 \;\;\;\;\;\; , \;\;\;\;\;\; \tau = t/m^{2}
\ee
and define the dilogarithm for arbitrary complex argument by
\be
\li (x) = - \int_{0}^{1} dt \,\frac{\ln (1 - x t)}{t} .
\ee
Then we have
\be
A(m)=1- \ln \frac{m^2}{\mu^2}
\ee
\be
B_{1}(s)  = 2 - \ln (\frac{-s-i \epsilon}{\mu^2})
\ee
\bear
    B_{2}(s)  = -\ln \frac{m^2}{\mu^2} - G (\sigma)
\eear
where
\be
 G(x) = \left\{   \begin{array}{lc}
2 \sqrt{\frac{4-x}{x}} \arcsin \frac{\sqrt{x}}{2}-2 &  0<x\leq 4
\\*[0.5cm]
2 \sqrt{\frac{x-4}{x}} (-i \frac{\pi}{2} +\acosh \frac{\sqrt{x}}{2} )
-2 &  x>4   \\*[0.5cm]
2 \sqrt{\frac{x-4}{x}} \asinh \frac{\sqrt{-x}}{2}-2 & x \leq 0
\end{array}.
\right.
\ee

The functions $C_1 (s), C_2 (s)$, arise from loop diagrams with three
internal propagators. Following ref.~\cite{kn:velt} they are given by
\be
C_{1}(s) = \frac{1}{\sigma}\,\left\{\li \left(1+i \epsilon +\sigma \right)
           -\frac{\pi^2}{6} \right\}
\ee
\bear
C_{2}(s) &  = & \frac{1}{\sigma} \left\{ \frac{\pi^2}{6}
         - \li \left(\frac{1-\sigma}{1-i \epsilon \sigma}\right)
         - \li \left(\frac{1}{1- \sigma l _{+} - i\epsilon} \right)
          + \right. \nonumber \\*[0.2cm]
 &  & \left. \li \left( \frac{1-\sigma}{1-\sigma l_{+}- i\epsilon}
     \right) -
    \li \left(\frac{1}{1-\sigma l_{-}+ i \epsilon} \right) +
     \li \left(\frac{1}{1-\sigma l_{-}+ i \epsilon} \right) \right\}
\eear
\newpage
where
\be
 l_{\pm}  = \frac{1}{2} \left(1 \pm\sqrt{1-4 /\sigma }\right) .
\ee
Finally the expression for the box diagram $D(s,t)$, following
ref.~\cite{kn:uli}, is given by
\bear
\lefteqn{(1+ \tau)\; (y - \frac{1}{y})\; D(s,t)  =
  2\;\li (1+\frac{z}{y}) -2\; \li (1+z y) + } \nonumber \\*[0.2cm]
  & & 4\; \li (1+y) + \; \ln^{2}\/(-y) \; - \;\frac{1}{2} \ln^{2} (-z\, y)
    \; + \;    2 \ln (-y) \ln (-  \tau -i \epsilon)
          \nonumber \\*[0.2cm]
    &  & \mbox{} - \eta (z,-y)\;  \ln (2+z y+ \frac{1}{z y}) \;
     - \; \eta (\frac{1}{z},-y) \; \ln (2+\frac{z}{y} +\frac{y}{z})
\eear
with
\[
z=1-\frac{\sigma}{2}+\left\{\frac{\sigma^{2}}{4} - \sigma
-i\epsilon \,\left( 2- \frac{\sigma}{2}\right) \right\}^{1/2}
\]

\[
y= \frac{\sigma \tau -2 \tau -2 +\sqrt{\sigma\tau (\sigma\tau -4 \tau
-4) + i \epsilon\tau (1 + \tau)}}{2 (1+ \tau)}
\]
where
\be
\eta (a,b) = \ln\,(a b)\; - \;\ln a\; - \;\ln b
\ee
as defined by 't Hooft and Veltman \cite{kn:velt}.

\newpage
 \newpage

\noindent
{\Large\bf Figure captions}
\vspace{0.5cm}

\begin{itemize}
\item
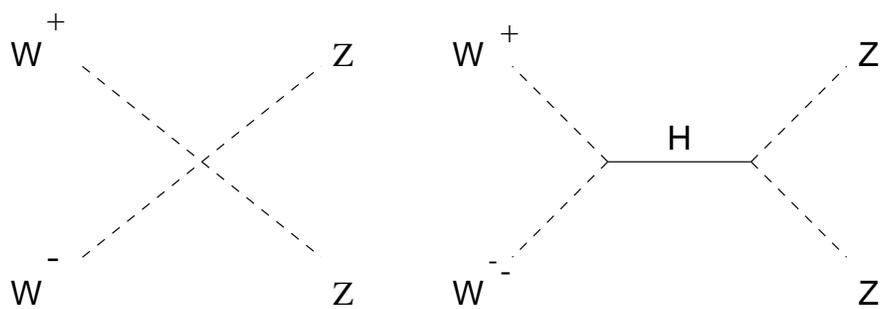
{\sl Figure 1} : The tree level Feynman diagrams contributing to  \wz.
\item
{\sl Figure 2a} : The one-loop  Higgs   Boson self-energy.
\item
{\sl Figure 2b} : The sum of diagrams that replaces the
      propagator on the resonance region.
\item
{\sl Figure 2c} : The Higgs   Boson self-energy contribution to \wz.  As
explained in section~3.2~A, for energies around the Higgs mass the
resummed  propagator is used from diagrams of Fig.~2b.
\item
{\sl Figure 3} : The one-loop gauge  Boson self-energy.
\item
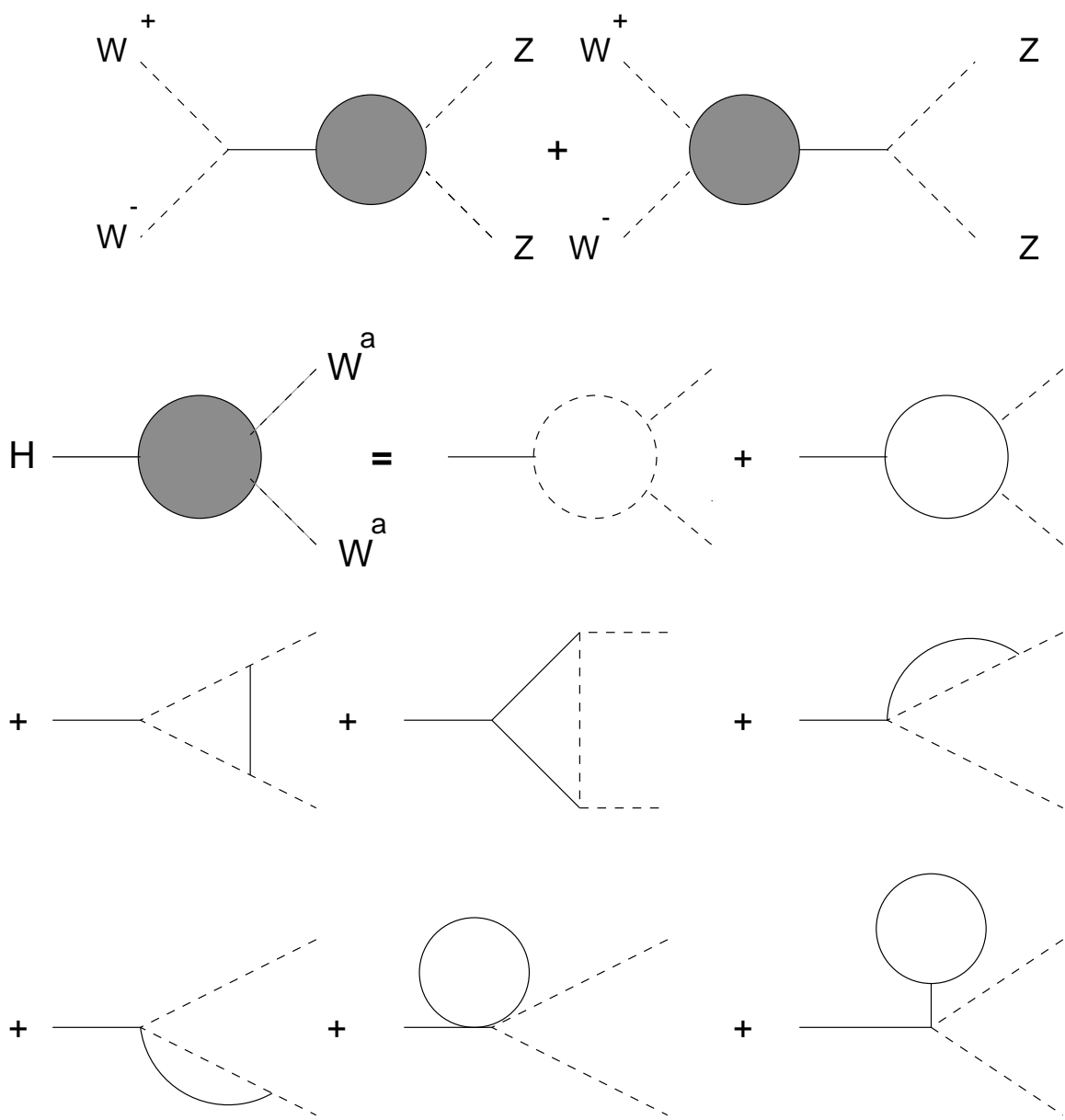
{\sl Figure 4} : The one-loop contribution to \wz\ scattering amplitude
through the $H w^{a} w^{a}$\ vertex.
\item
{\sl Figure 5a} : The bubble diagrams that contribute to \wz.
\item
{\sl Figure 5b} : The rest of the 1-PI diagrams with two internal
propagators.
\item
{\sl Figure 5c-d} : The Feynman diagrams with three internal propagators.
\item
{\sl Figure 5e} : The Feynman diagrams with four internal propagators.
\item
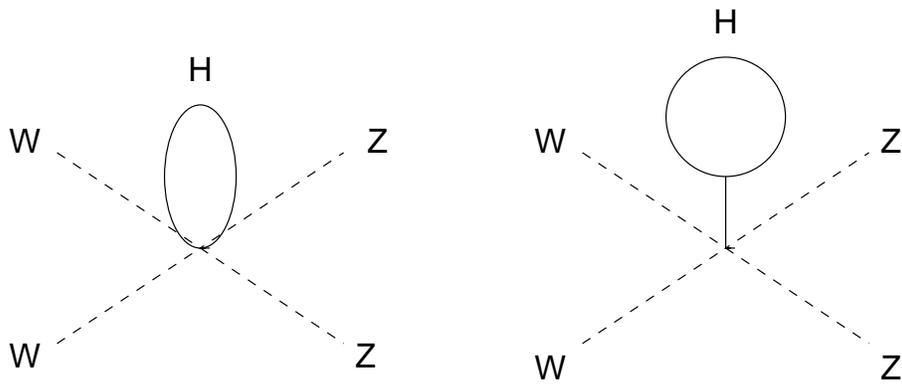
{\sl Figure 6} : The tadpole diagrams.
\item
{\sl Figure 7a} : The total cross section for $W^{+}_{L} W^{-}_{L}
\rightarrow W^{+}_{L} W^{-}_{L}$, in the $SU(4)/Sp(4)$ with a Higgs mass
$m_{H} = 718$ GeV, and using the values in eq.~(\ref{eq:vl}) as a
function of  $s$. Solid lines correspond to tree level and dashed lines
to one-loop.
\item
{\sl Figure 7b} : The total cross section for $W^{+}_{L} W^{-}_{L}
\rightarrow W^{+}_{L} W^{-}_{L}$ in the Standard Model with a Higgs mass
$m_{H} = 718$ GeV, as a function of  $s$. Solid lines
correspond to tree level and  dashed lines to one-loop.
\item
{\sl Figure 8a} : The total Higgs decay width as a function of the Higgs
 mass for the $SU(4)/Sp\/(4)$ model with the choise in
 eq.~(\ref{eq:vl}). Solid lines correspond to tree level and
dashed lines to one-loop.
\item
{\sl Figure 8b} : The total Higgs decay width as a function of the Higgs
 mass for the standard model.  Solid lines  correspond to tree level
 and dashed lines  to one-loop.
\end{itemize}


\begin{thebibliography}{99}
\bibitem{kn:thacker}B.W. Lee, C. Quigg and H. Thacker, \prd{\bf 16}
 (1977) 1519.
\bibitem{kn:thooft}G.'t Hooft, in: Recent Developments in Gauge
Theories, eds., G.~'t~Hooft (Plenum Press, New York, 1980).
\bibitem{kn:phi4} K.~G.~Wilson, \pr {\bf B 4} (1971) 3184;\\
 K.~G.~Wilson and J.~Kogut, \prp {\bf 12} (1974) 76;\\
 G.~A.~Baker and J.~M.~Kincaid, \jsp {\bf 24} (1981) 469;\\
 M.~Aizenmann, \prl {\bf 47} (1981) 1; \cmp {\bf 86} (1982) 1;\\
 J.~Fr\"{o}hlich, \np{\bf 200} [FS4] (1982) 281;\\
 A.~D.~Sokal, Ann. Inst. H. Poincar\'{e} {\bf A 37} (1982) 317;\\
 C.~Arag\~{a}o de Carvalho, S.~Caracciolo and J.~Fr\"{o}hlich, \np{\bf
 215} (1983) 209;\\
 B.~Freedmann, P.~Smolensky and D.~Weingarten, \pl{\bf 113} (1982) 481;\\
 for a review, see D.~J.~E.~Callaway, Phys. Rep. {\bf 167} (1988) 241.
\bibitem{kn:o(n)} M.~L\"{u}scher and P.~Weisz, \np{\bf 318} (1989) 705;\\
 J.~Kuti, L.~Lin and Y.~Shen, \prl {\bf 61} (1988) 678;\\
 A.~Hasenfratz, K.~Jansen, C.~B.~Lang, T.~Neuhaus and H.~Yoneyama, \pl
 {\bf 199} (1987) 531;\\
 A.~Hasenfratz, K.~Jansen, J.~Jers\'{a}k, C.~B.~Lang, T.~Neuhaus and
 H.~Yoneyama, \np{\bf 317} (1989) 81;\\
 G.~Bhanot, K.~Bitar, U.~M.~Heller and H.~Neuberger, \np{\bf 353} (1991) 551.
\bibitem{kn:dn} R.~Dashen and H.~Neuberger, \prl {\bf 50} (1983) 1897.
\bibitem{kn:ehlq}See, for example E.~J.~Eichten, C.~Quigg, I.~Hinchliffe
          and K.~D.~Lane, \rmp{\bf 56} (1984) 579.
\bibitem{kn:cahn}R. N. Cahn and S. Dawson, \pl{\bf 136} (1984) 196.
\bibitem{kn:chanowitz} M. Chanowitz and M. K. Gaillard, \np{\bf 261}
    (1985) 379.
\bibitem{kn:ET} J. Cornwall, D. Levin and G. Tiktopoulos, \prd{\bf 10},
  (1974) 1145;\\ C. Vayonakis, Lett.\ Nuovo Cimento 17 (1976) 383; \\
 G. Gounaris, R. K\"{o}gerler and H. Neufeld, \prd{\bf 34} (1986)3257;\\
 J. Bagger and C. Schmidt, \prd{\bf 41} (1990) 264;\\
 H. Veltman,  \prd{\bf 41} (1990) 2;\\
 W. B. Kilgore, \pl{\bf 294} (1992) 257;\\
 H. J. He and Y. P. Kuang, \prl{\bf 69} (1992) 2619.
\bibitem{kn:ccwz}S. Coleman, J. Wess and B. Zumino, \pr {\bf 177}
   (1969) 2239;\\ C.~G.~Callan, S.~Coleman, J.~Wess and B.~Zumino,
               \pr {\bf 177} (1969) 2247.
\bibitem{kn:gm} S. Weinberg, Physica 96 A (1979) 327;\\
            H. Georgi and A. Manohar, \np{\bf 234} (1984) 189;\\
            H. Georgi, \pl{\bf 298} (1993) 187.
\bibitem{kn:bible}For a review, see  H. Georgi, {\sl Weak Interactions and
    Modern Particle Theory} (Benjamin/Cummings, Menlo Park C.A. 1984).
\bibitem{kn:short}R. S. Chivukula and V. Koulovassilopoulos, \pl{\bf 309}
                  (1993) 371.
\bibitem{kn:gk}D. B. Kaplan and H. Georgi, \pl{\bf 136} (1984) 183.
\bibitem{kn:dgk}D. B. Kaplan, S. Dimopoulos and H. Georgi, \pl{\bf 136}
                (1984) 187;\\
        H. Georgi, D. B. Kaplan and P. Galison \pl{\bf 143} (1984) 152;\\
        H. Georgi and D. B. Kaplan, \pl{\bf 145} (1984) 216;\\
     M. J. Dugan, H. Georgi and D. B. Kaplan, \np{\bf 254} (1985) 299.
\bibitem{kn:witten}E. Witten, \pl{\bf 117} (1982) 324.
\bibitem{kn:preskill} M.E. Peskin, \np{\bf 175} (1980) 197;\\
                      J. Preskill, \np{\bf 177} (1981) 21.
\bibitem{kn:kosower}D. Kosower, \pl{\bf 144} (1984) 215;\\
                    C. Vafa and E. Witten, \np{\bf 234} (1984) 1.
\bibitem{kn:leshouches}H. Georgi, {\sl $SU(2)\times U(1)$ Breaking,
                   Compositeness and Flavor}, in Les Houches (1985).
\bibitem{kn:salam}S. Kamefuchi, L. O' Raifeartaigh and A. Salam Nucl.
                  Phys. {\bf 28} (1961) 529;\\
                 R. Haag, \pr {\bf 112}, (1958) 669;\\
                 D. Ruelle, Helv. Phys. Acta {\bf 35}, (1962) 34;\\
                 H. J. Borchers, Nuovo Cimento {\bf 25}, (1960) 270.
\bibitem{kn:cara} T. Appelquist and J. Carazzone, \prd{\bf 11}
                 (1975) 2856.
\bibitem{kn:gl}J. Gasser and H. Leutwyler, \ap  158 (1984) 142;
                \np{\bf 250} (1985) 465.
\bibitem{kn:samuel}E. H. Simmons, \np{\bf 312} (1989) 253;\\
  S. Samuel, \np{\bf 347} (1990) 625;\\
  A. Kagan and S. Samuel, Int.\ J.\ Mod.\ Phys. {\bf A7} (1992) 1123;\\
  C. D. Carone and E. H. Simmons, \np{\bf 397} (1993) 591;\\
  C. D. Carone and H. Georgi,  HUTP-93-A015 , hep-ph@xxx.lanl.gov -
   9308205 (1993).
\bibitem{kn:golden}M. S. Chanowitz, H. Georgi, M. Golden, \prd {\bf 36}
            (1987) 1490;  \prl {\bf 57}(1986) 2344.
\bibitem{kn:dobado}M. Golden, in: {\sl Beyond the standard model} (Iowa
  State University, 1988),\\ eds. K.~Whisnant and B.-L~Young (World
  Scientific, Singapore, 1989) p.~111;\\
  A. Dobado and M. Herrero, \pl {\bf 228} (1989) 495;  (1989) 505.
\bibitem{kn:donoghue} J.~F. Donoghue and C. Ramirez, \pl {\bf 234}
  (1990)361;\\   S. Dawson and G. Valencia, \np{\bf 352} (1991) 27.
\bibitem{kn:ecker} G.~Ecker, A.~Pich and E.~de Rafael, \np{\bf 321}
        (1989) 311;\\ J.~F. Donoghue, C. Ramirez and G. Valencia,
         \prd{\bf 38} (1988) 2195; \prd {\bf 39} (1989) 1947.
\bibitem{kn:pagels} L.-F. Li and H. Pagels, \prl{\bf 26} (1971) 1089;\\
   see also H. Pagels, \prp {\bf 16C} (1975) 221.
\bibitem{kn:dawil} S. Dawson and S. D. Willenbrock, \prd{\bf 40} (1989) 2880;
   \prl {\bf 62} (1989) 1232.
\bibitem{kn:valencia92}G. Valencia and S.~D.~Willenbrock, \prd {\bf 46}
                       (1992) 2247.
\bibitem{kn:durand} L. Durand, J. M. Johnson and J. L. Lopez, \prd {\bf 45}
                   (1992) 3112.
\bibitem{kn:marciano}W.~J.~Marciano and S.~D.~Willenbrock, \prd{\bf 37},
                     (1988) 2509.
\bibitem{kn:valencia}S.~D.~Willenbrock and G. Valencia, \prd {\bf 42}
                     (1990) 853.
\bibitem{kn:shape} S.~D. Willenbrock and G. Valencia, \pl{\bf 247} (1990) 341.
\bibitem{kn:dawsontop}S. Dawson and G. Valencia, \np{\bf 348} (1991) 23.
\bibitem{kn:velt}G.~`t~Hooft and M.~Veltman, \np{\bf 153} (1979) 365.
\bibitem{kn:uli} A. Denner, U. Nierste and R. Scharf, \np{\bf 367} (1991)
   637;\\ U. Nierste, Diplomarbeit, W\"{u}rzburg (1991).
\end{thebibliography}
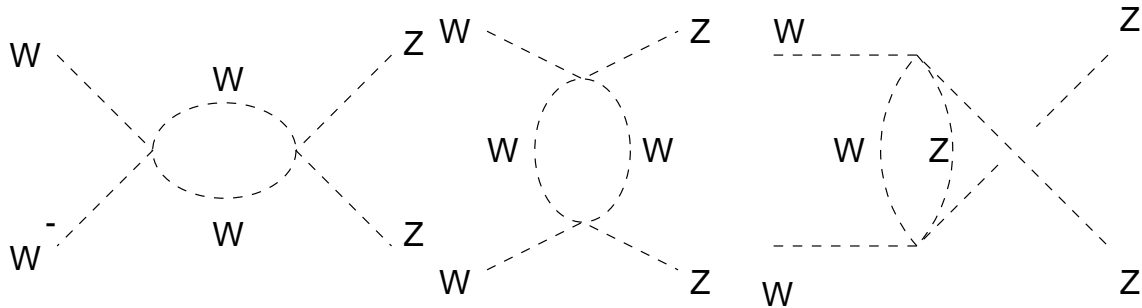
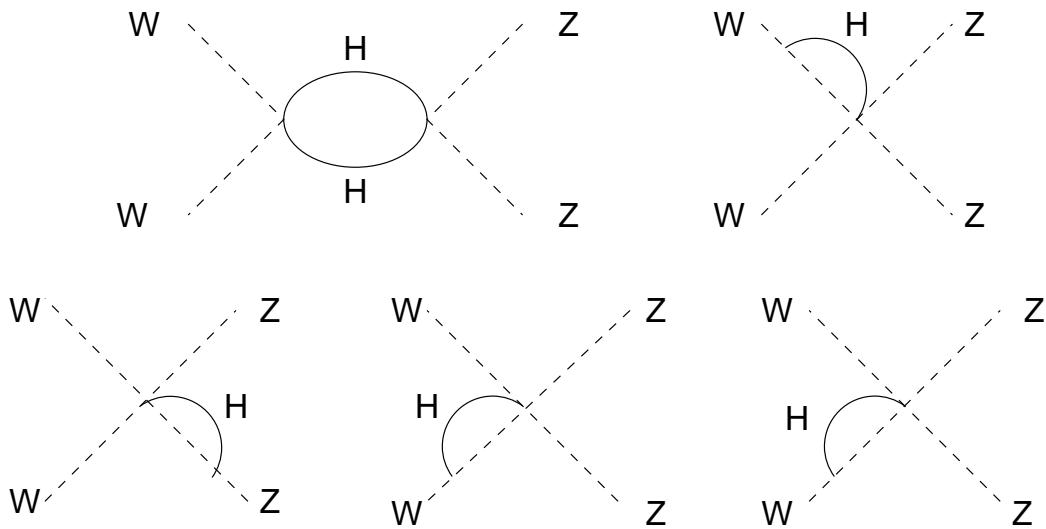
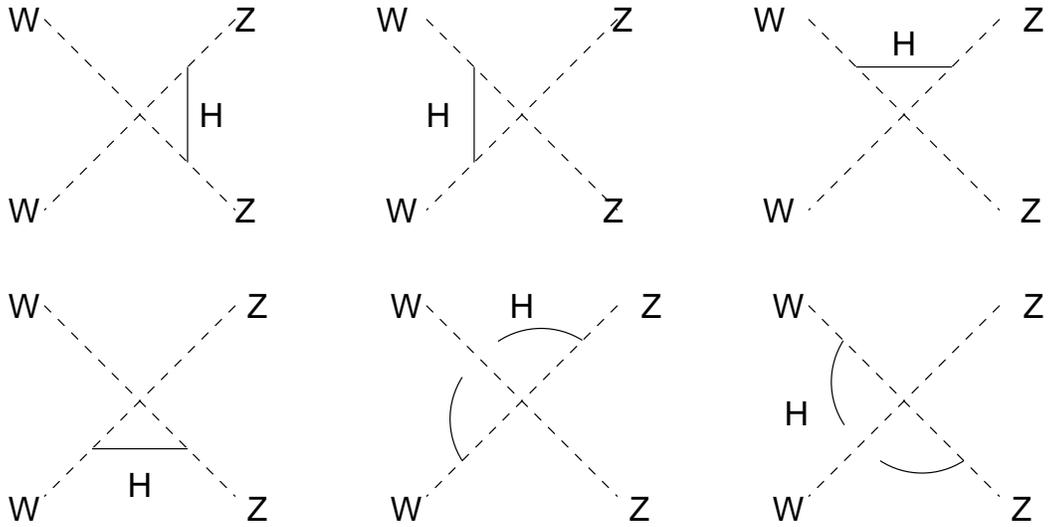
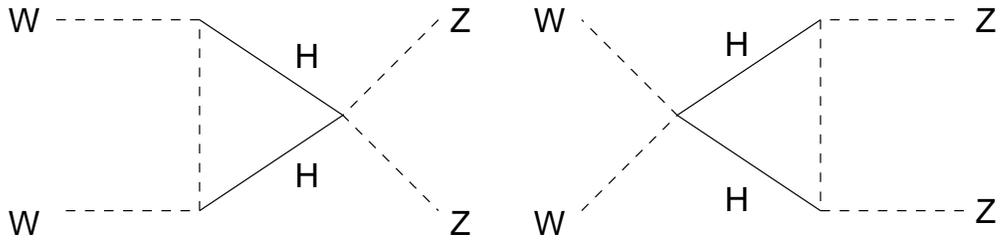
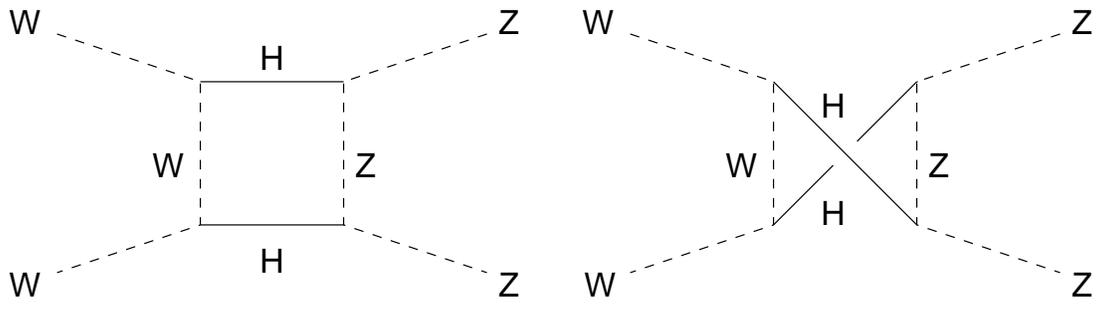
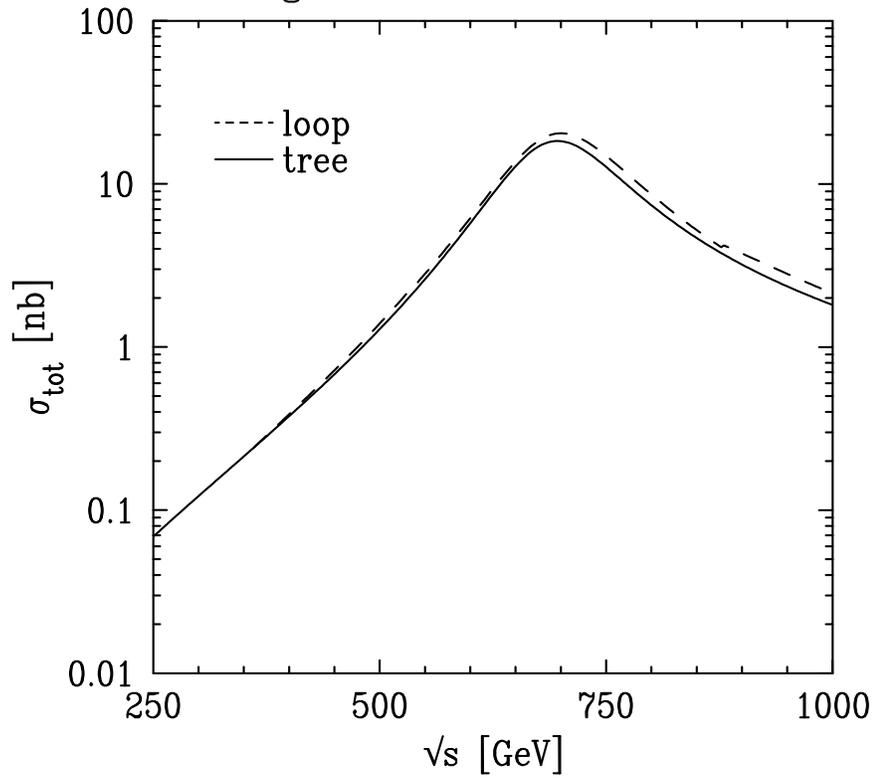
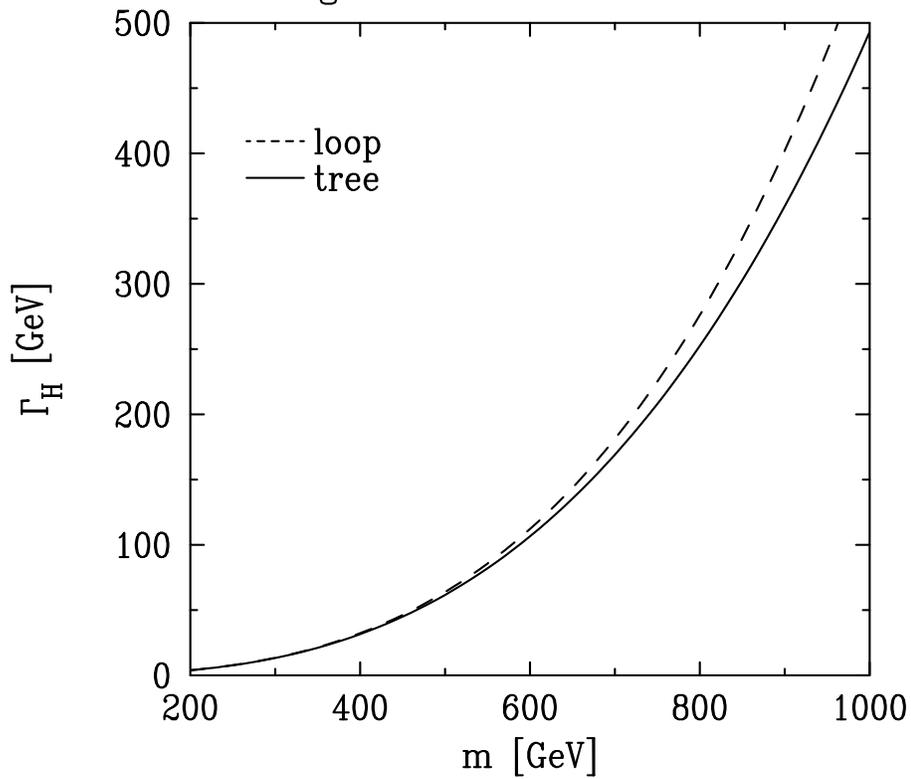
\end{document}